%% file: spa_tvcg.tex
\documentclass[10pt,journal,compsoc]{IEEEtran}
\ifCLASSOPTIONcompsoc
  \usepackage[nocompress]{cite}
\else
  \usepackage{cite}
\fi
\ifCLASSINFOpdf
  \usepackage[pdftex]{graphicx}
\else
\fi
\usepackage{amsmath}

\usepackage{verbatim}

\usepackage{color}
\newcommand{\purple}{\textcolor[RGB]{0,0,0}}

\usepackage[normalem]{ulem}

\begin{document}
%
\title{Example-based Real-time Clothing Synthesis for Virtual Agents}

\author{Nannan Wu,
Qianwen Chao,~\IEEEmembership{Member,~IEEE,}
Yanzhen Chen,
Weiwei Xu,~\IEEEmembership{Member,~IEEE,}\\
Chen Liu,
Dinesh Manocha,~\IEEEmembership{Fellow,~IEEE,}
Wenxin Sun,
Yi Han,
Xinran Yao,
Xiaogang Jin*,~\IEEEmembership{Member,~IEEE}
\IEEEcompsocitemizethanks{\IEEEcompsocthanksitem N. Wu, Y. Chen, W. Xu, W. Sun, Y. Han, X. Yao and X. Jin are with State Key Lab of CAD\&CG, Zhejiang University, Hangzhou 310058, P. R. China. E-mail: \{wunannanzju, choconutscyz\}@gmail.com, xww@cad.zju.edu.cn, 3170101802@zju.edu.cn, hany28@mail2.sysu.edu.cn, yaoxinran1022@163.com, jin@cad.zju.edu.cn.
\IEEEcompsocthanksitem Q. Chao is with the Department of Computer Science, Xidian University, Xi’an 710038, P. R. China. E-mail: chaoqianwen15@gmail.com.
\IEEEcompsocthanksitem C. Liu is with Zhejiang Linctex Digital Technology Co. Ltd., China. E-mail                              : eric.liu@linctex.com. 
\IEEEcompsocthanksitem Dinesh Manocha is at University of Maryland, USA. E-mail                                                : dm@cs.umd.edu.
\IEEEcompsocthanksitem X. Jin is the corresponding author.
}
}


\markboth{Example-based Real-Time Clothing Synthesis for Virtual Agents}%
{}

%



\IEEEtitleabstractindextext{%
\begin{abstract}
We present a real-time cloth animation method for dressing virtual humans of \purple{various shapes and poses}. Our approach formulates the clothing deformation as a high-dimensional function of body shape parameters and pose parameters. 
In order to accelerate the computation, our formulation factorizes the clothing deformation into two independent components: the deformation introduced by body pose variation (Clothing Pose Model) and the deformation from body shape variation (Clothing Shape Model).
Furthermore, we sample and cluster the poses spanning the entire pose space and use those clusters to efficiently calculate the anchoring points.
We also introduce a sensitivity-based distance measurement to both find nearby anchoring points and evaluate their contributions to the final animation. Given a query shape and pose of the virtual agent, we synthesize the resulting clothing deformation by blending the Taylor expansion results of nearby anchoring points. Compared to previous methods, our approach is general and able to add the shape dimension to any clothing pose model. 
Furthermore, we can animate clothing represented with tens of thousands of vertices at 50+ FPS on a CPU.
We also conduct a user evaluation  and show that our method can improve a user's perception of dressed  virtual agents in an immersive virtual environment compared to a conventional linear blend skinning method.%
\end{abstract}

\begin{IEEEkeywords}
clothing animation, virtual agents, social VR,  virtual try on clothing shape models.
\end{IEEEkeywords}}

\maketitle

\begin{figure*}
   \includegraphics[width=\linewidth]{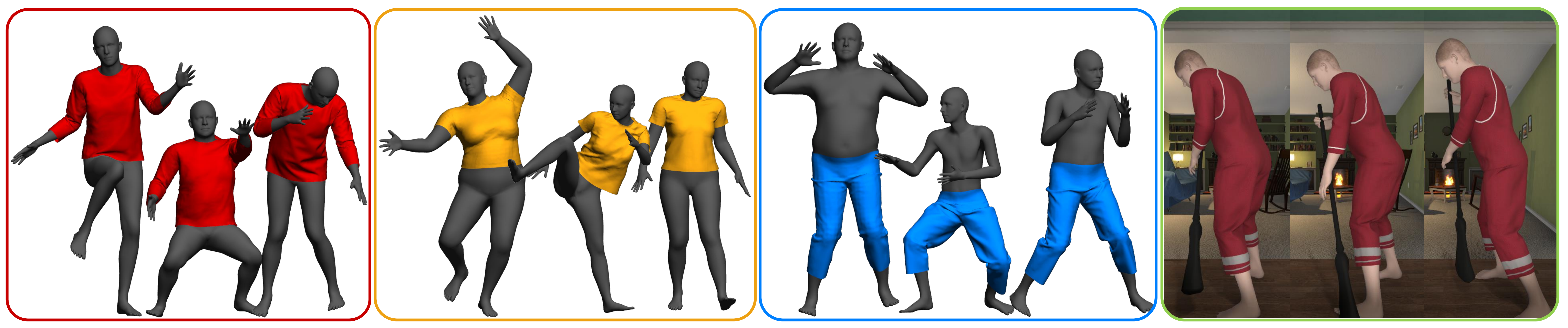}
   \centering
   \caption{Our method can generate real-time clothing animation results with detailed wrinkles for various body poses and shapes with different clothing types on a commodity CPU. Our method can also be applied in VR scenarios.}
   \label{fig:teaser}
\end{figure*}

\IEEEdisplaynontitleabstractindextext

%
\IEEEpeerreviewmaketitle

\input{spa_body}


%

%

\ifCLASSOPTIONcompsoc
  \section*{Acknowledgments}
\else
  \section*{Acknowledgment}
\fi

Xiaogang Jin was supported by the National Key R\&D Program of China (Grant No. 2017YFB1002600), the Ningbo Major Special Projects of the “Science and Technology Innovation 2025” (Grant No. 2020Z007), the National Natural Science Foundation of China (Grant Nos. 61732015, 61972344), and the Key Research and Development Program of Zhejiang Province (Grant No. 2020C03096). Qianwen Chao was supported by the National Natural Science Foundation of China (Grant No. 61702393).
We thank Zhejiang Linctex Digital Technology Co. Ltd for the help on designing the female T-shirt in Figure \ref{fig:moreresult}.

\ifCLASSOPTIONcaptionsoff
  \newpage
\fi




\bibliographystyle{IEEEtran}
\bibliography{spa}

%
%
%

%

\begin{IEEEbiography}[{\includegraphics[width=1in,height=1.25in,clip,keepaspectratio]{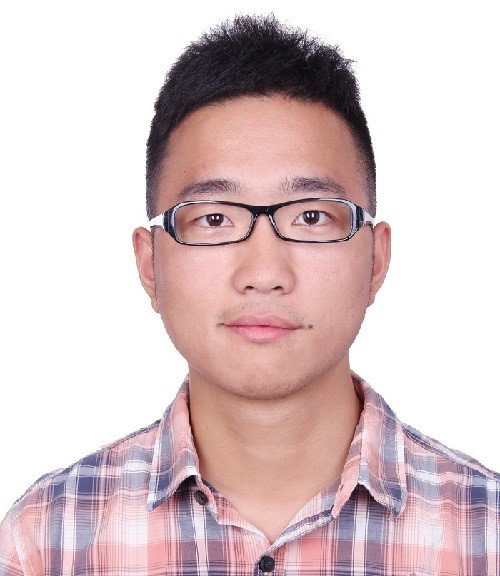}}]{Nannan Wu} received his BSc degree in Computer Science from Nanjing University of Science and Technology, China, in 2013. He is currently a PhD candidate in the State Key Lab of CAD\&CG, Zhejiang University, China. His main research interests include cloth simulation, virtual try-on and computer animation.
\end{IEEEbiography}

\vspace{-10mm}
\begin{IEEEbiography}[{\includegraphics[width=1in,height=1.25in,clip,keepaspectratio]{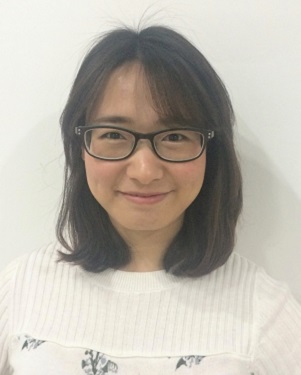}}]{Qianwen Chao} is a Lecturer of Computer Science at Xidian University, China. She earned her Ph.D. degree in Computer Science from the State Key Laboratory of CAD\&CG, Zhejiang University in 2016. Prior that, she received her B.S. degree in computer science in 2011 from Xidian University. Her main research interests include crowd animation and swarm micro-robotics.
\end{IEEEbiography}

\vspace{-10mm}
\begin{IEEEbiography}[{\includegraphics[width=1in,height=1.25in,clip,keepaspectratio]{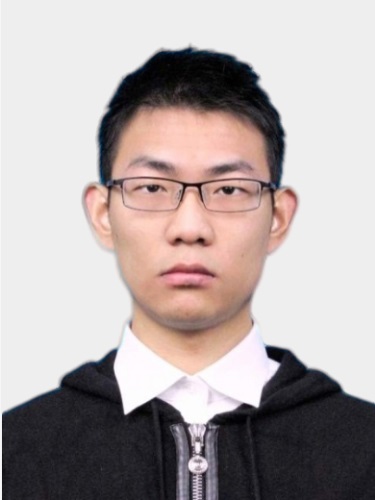}}]{Yanzhen Chen} received the BSc degree in digital media technology from Zhejiang University, P. R. China, in 2020. He is currently working toward the MSc degree at Zhejiang University. His research interests include computer graphics and computer animation.
\end{IEEEbiography}

\vspace{-10mm}
\begin{IEEEbiography}[{\includegraphics[width=1in,height=1.25in,clip,keepaspectratio]{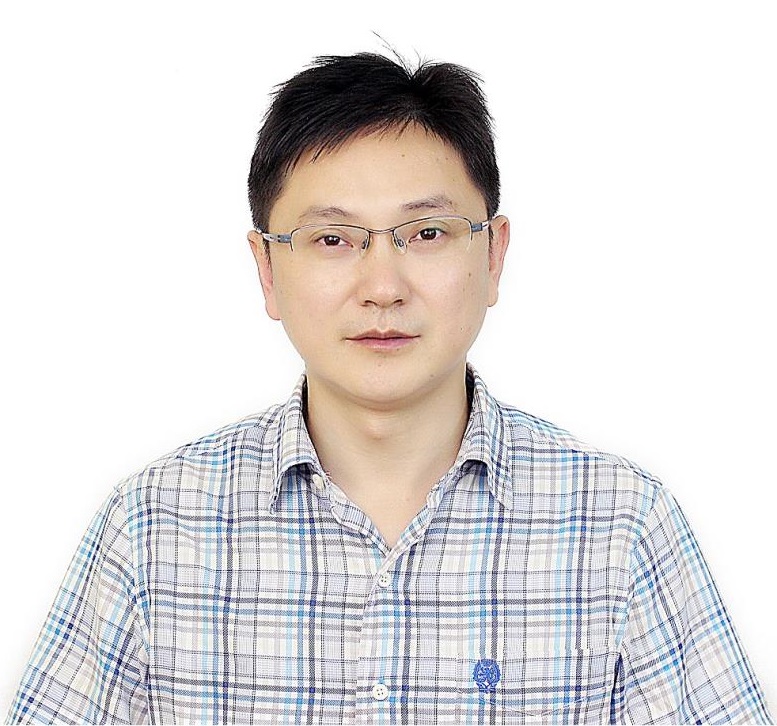}}]{Weiwei Xu} is a researcher with the State Key Lab of CAD \& CG, College of Computer Science, Zhejiang University, awardee of NSFC Excellent Young Scholars Program in 2013. His mainresearch interests include the digital geometry processing, physical simulation, computer vision and virtual reality. He has published around 70 papers on international graphics journals and conferences, including 19 papers on ACM TOG. He is a member of the IEEE.
\end{IEEEbiography}

\vspace{-10mm}
\begin{IEEEbiography}[{\includegraphics[width=1in,height=1.25in,clip,keepaspectratio]{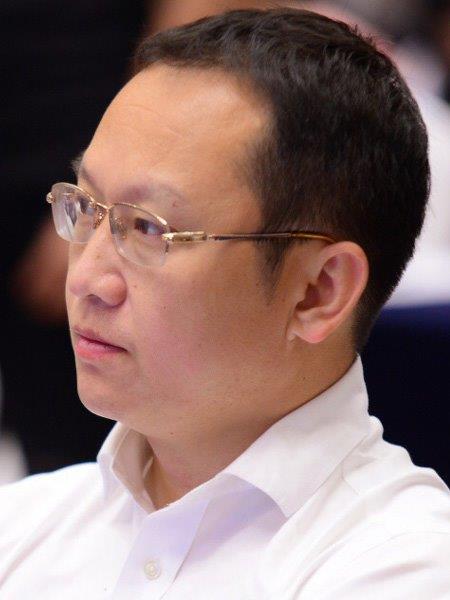}}]{Chen Liu} is the CEO of Zhejiang Linctex Digital Technology Co. Ltd., China. He received his BSc degree in 1993 from Zhejiang Institute of Silk Technology, MSc degree in MBA in 2000 from Zhejiang University, and another MSc degree in applied computer science in 2003 from Vrije Universiteit Brussel (VUB). His current research interests mainly focus on computer-aided clothing design, cloth animation, digital face modeling, and new retail.
\end{IEEEbiography}

\vspace{-10mm}
\begin{IEEEbiography}[{\includegraphics[width=1in,height=1.25in,clip,keepaspectratio]{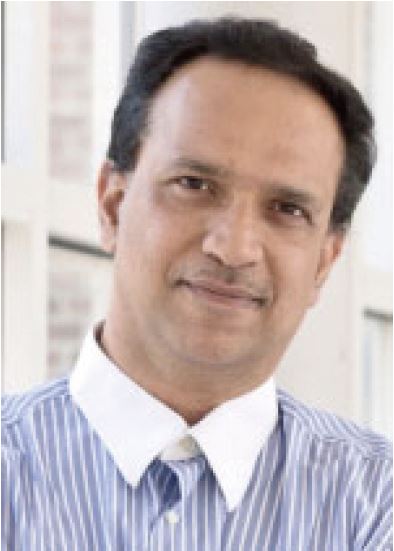}}]{Dinesh Manocha} is the Paul Chrisman Iribe Chair in Computer Science \& Electrical and Computer Engineering at the University of Maryland College Park. He is also the Phi Delta Theta/Matthew Mason Distinguished Professor Emeritus of Computer Science at the University
of North Carolina - Chapel Hill. He has won many awards, including Alfred P. Sloan Research Fellow, the NSF Career Award, the
ONR Young Investigator Award, and the Hettleman Prize for scholarly achievement. His research interests include multi-agent simulation, virtual environments, physicallybased modeling, and robotics. His group has developed a number of packages for multi-agent simulation, crowd simulation, and physicsbased simulation that have been used by hundreds of thousands of users and licensed to more than 60 commercial vendors. He has published more than 480 papers and supervised more than 35 PhD dissertations. He is an inventor of 9 patents, several of which have been licensed to industry. His work has been covered by the New York Times, NPR, Boston Globe,Washington Post, ZDNet, as well as DARPA Legacy Press Release. He is a Fellow of AAAI, AAAS, ACM, and IEEE and also received the Distinguished Alumni Award from IIT Delhi. See http://www.cs.umd.edu/people/dmanocha.
\end{IEEEbiography}

\vspace{-10mm}
\begin{IEEEbiography}[{\includegraphics[width=1in,height=1.25in,clip,keepaspectratio]{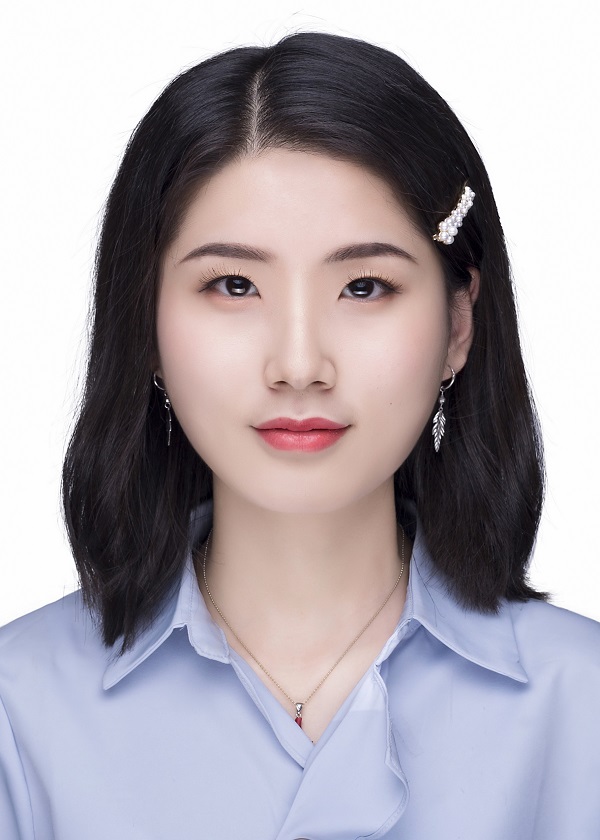}}]{Wenxin Sun} is currently an undergraduate student in Zhejiang University majoring in digital media technology. Her main research interests include image processing and facial animation.
\end{IEEEbiography}

\vspace{-10mm}
\begin{IEEEbiography}[{\includegraphics[width=1in,height=1.25in,clip,keepaspectratio]{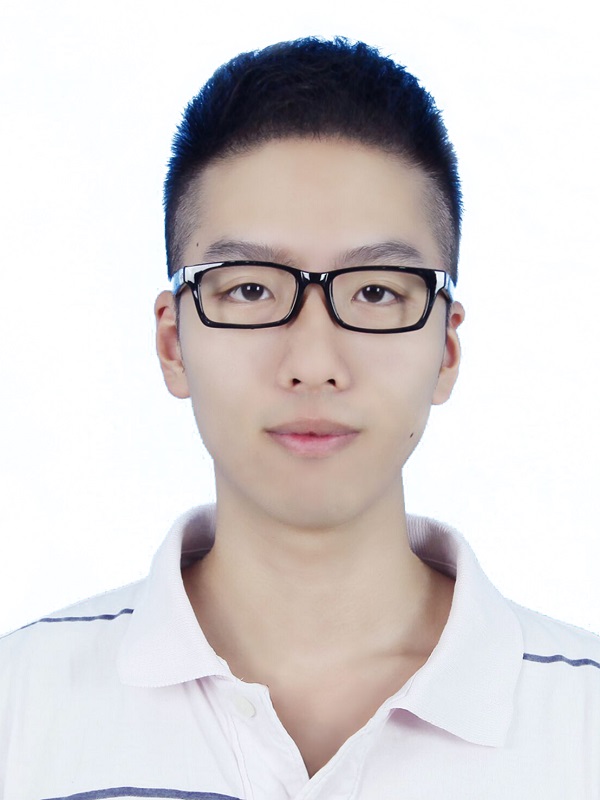}}]{Yi Han} is a Ph.D. candidate in the State Key Lab of CAD \& CG, Zhejiang University. Prior to that, he received his bachelor's degree in software engineering in 2017 from Sun Yat-sen University. Recently, his main researches focus on crowd animation, traffic simulation, and computer animation.
\end{IEEEbiography}

\vspace{-10mm}
\begin{IEEEbiography}[{\includegraphics[width=1in,height=1.25in,clip,keepaspectratio]{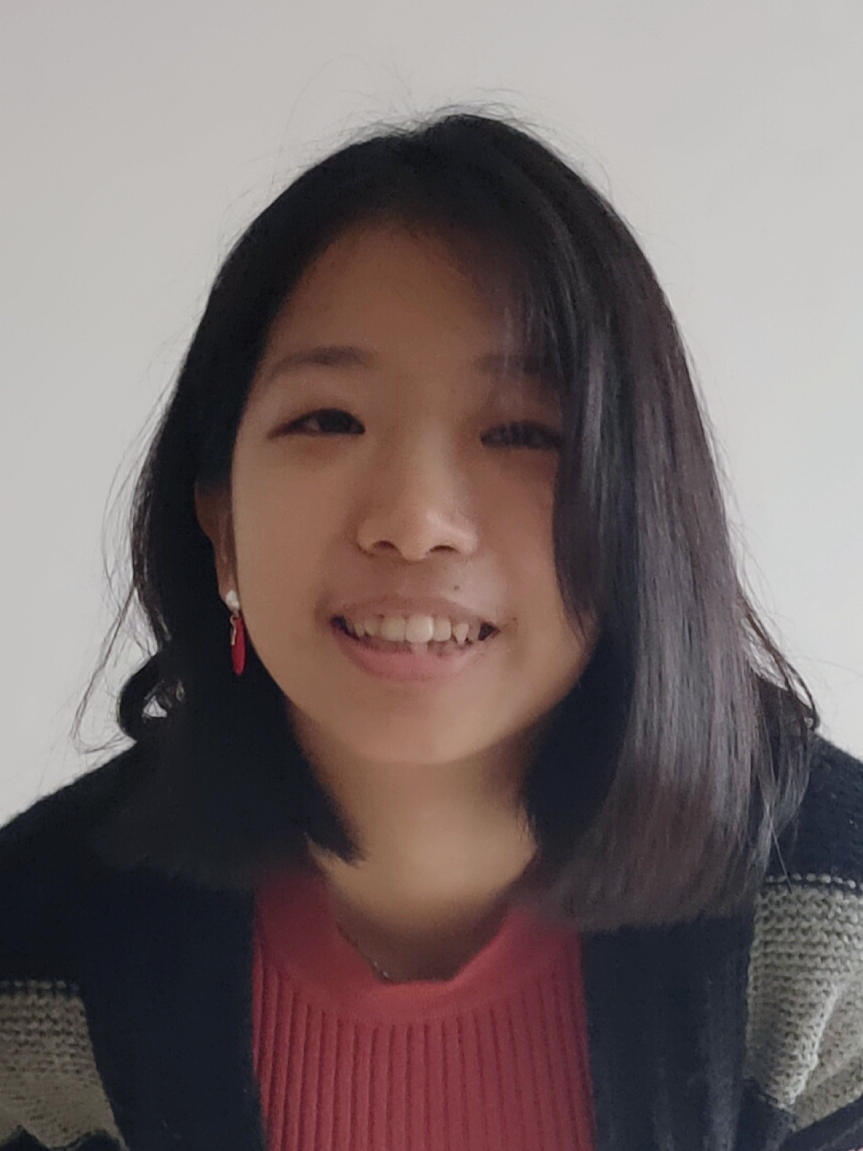}}]{Xinran Yao} is a master candidate in the State Key Lab of CAD\&CG, Zhejiang University, China. She graduated from Zhejiang University with a bachelor's degree in 2018 majoring in digital media technology. In recent years, her main researches focus on crowd animation and game AI.
\end{IEEEbiography}

\vspace{-10mm}
\begin{IEEEbiography}[{\includegraphics[width=1in,height=1.25in,clip,keepaspectratio]{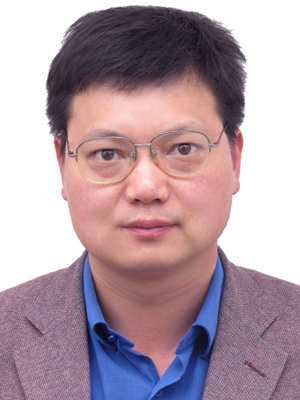}}]{Xiaogang Jin} is a professor of the State Key Lab of CAD\&CG, Zhejiang University, China. He received his BSc degree in computer science in 1989, MSc and PhD degrees in applied mathematics in 1992 and 1995, all from Zhejiang University. His current research interests include traffic simulation, insect swarm simulation, physically based animation, cloth animation, special effects simulation, implicit surface computing, non-photorealistic rendering, computer generated marbling, and digital geometry processing. He received an ACM Recognition of Service Award in 2015, and the Best Paper Awards from CASA 2017 and CASA 2018. He is a member of the IEEE and ACM.
\end{IEEEbiography}






\end{document}

%% file: spa_body.tex
\section{Introduction}
\label{sec:introduction}
There is considerable interest in generating human-like virtual agents for AR and VR applications.
These agents are used to generate immersive social experiences for games, training, entertainment, virtual space visitations, or social-phobia treatments. In order to maintain the sense of presence in virtual environments, it is \purple{essential} that these virtual agents look realistic and interact in a plausible manner~\cite{bailenson2005independent}.

There has been a great deal of work on improving the realism of virtual humans in terms of rendering, body shapes, facial expressions, hair, locomotion, etc. A key issue is related to dressing three-dimensional virtual humans using garments and animating  the cloth deformation corresponding to  draping and wrinkles. This is \purple{crucial} because up to $80$\% of a human body can be covered by clothing. As virtual agents move, bend, or interact with the environment, the clothing folds, wrinkles, and stretches to conform to the virtual agents' poses. Moreover, the garments may stick to themselves and/or other pieces of clothing.
This problem is also important in the context of efficient virtual try-on simulation. In these applications, the user can experience the fit of several garments  on their 3D models or avatars.  The goal is to experience real-life clothes on virtual models and evaluate the garment behavior when a user moves around or stands in different poses. Recently, many AR-based interfaces have been proposed for virtual try-on.  All these applications must be able to simulate or animate the cloth and garments at interactive rates.



Physics-based simulation methods directly model the non-linear behavior of clothing and contacts to generate realistic cloth simulations~\cite{baraff1998large,narain2012adaptive,cirio2014yarn}. However,  high-resolution clothing meshes and \purple{ computationally} expensive nonlinear solvers are frequently used, making it difficult to simulate clothing at interactive rates (i.e. 30fps or more).
Moreover, dressing bodies of different shapes requires a separate simulation or setup for each body shape. This makes it difficult to use physics-based methods for real-time VR or virtual try-on applications. 

\purple{ Most of the data-driven clothing animation methods} synthesize the cloth deformation from pre-computed clothing deformation samples for different body poses~\cite{kim2013near,xu2014sensitivity}. In order to achieve real-time performance, these methods compromise the animation quality by simplifying the relationship between the deformed clothing and the underlying body poses by assuming linearity or locality. To consider the effect of body shape on cloth deformation, Peng et al.~\cite{guan2012drape} model the cloth deformation as a function of body motion and shape and resize the garment to fit the body. Recently, \cite{santesteban2019learning} presented a learning-based model to predict the deformation of a fixed-size garment under various poses and shapes. However, it uses an exhaustive database construction procedure that requires substantial computational resources \purple{offline}.



\noindent {\bf Main Results:}
In this paper, we present a novel real-time algorithm for cloth synthesis for VR and virtual try-on applications. We use a data-driven approach and present a new method to simulate plausible deformation effects at an interactive rate.  We factorize the clothing deformation into two independent components: the pose-dependent deformation and the shape-dependent deformation. We refer to the pose-dependent deformation as the {\em clothing pose model}, which is used to predict clothing deformations under various poses; we refer to the shape-dependent deformation as the {\em clothing shape model}, which can predict clothing deformations under various shapes. We also present a clothing synthesis scheme that combines these two components through Taylor expansion. Our method adopts a pose- and shape-dependent skinning scheme for clothing synthesis to meet the needs of real-time virtual try-on on a CPU for synthetic bodies of \purple{various shapes and poses} (as shown in Figure~\ref{fig:teaser}). The three novel components of our work are:

{\textbf{Clothing shape model:}}
\purple{We present a novel pose-independent clothing shape model.}
Given a set of clothing instances simulated on bodies with a specific pose and various shapes, we first reduce the \purple{dimensionality} with Principal Component Analysis (PCA). We use the first $k$ principal component coefficients spanning the clothing deformation space \purple{(we experimentally set $k=5$, as described in Section \ref{subsec:clothingshapemodel})}. Next, we map the body shape parameters to the coefficients of the PCA basis of the clothing deformation space. Therefore, given a set of body shape parameters, we can predict the corresponding clothing deformation result.  

{\textbf{Taylor expansion to approximate clothing deformations:}}
\purple{We present a real-time clothing synthesis method by considering both the pose-dependent deformation and the shape-dependent deformation using Taylor expansions.}
We represent clothing deformation as a function of body shape parameters $\beta$ and body pose parameters $\theta$, $f(\beta,\theta)$. \purple{The clothing deformation in the neighborhood of the given anchoring point $(\beta_0,\theta_0)$ is approximated with a Taylor expansion. The partial derivatives of the clothing deformation} are calculated numerically using the clothing pose model and the clothing shape model to predict $f(\beta_0, \theta_0+\Delta \theta)$ and $f(\beta_0+\Delta \beta, \theta_0)$, respectively. Given the new parameters $(\beta=\beta_0+\Delta \beta,\theta=\theta_0+\Delta \theta)$, we synthesize the clothing deformation by blending the Taylor expansion results from \purple{the} nearby anchoring points. \purple{Accordingly, our approach can add the shape dimension to any clothing pose model; this is in contrast to \cite{xu2014sensitivity}, which can only deal with pose-dependent deformations.} Moreover, we use a sensitivity-based measurement to find nearby anchoring points and to calculate the blending weights.

{\textbf{Sampling scheme:}}
\purple{We present a pose space analysis method to generate a compact example database in parallel, which can significantly reduce the offline computational cost.}
In order to generate plausible results,
we cluster the pose space into a small number of clusters. We use the cluster centers to calculate the anchoring points and thereby build a compact database. In our implementation, we only sample data points along the $\theta$ axis, which enables us to not have to train the input clothing pose model at various shapes. This sampling scheme can also be used to generate the database of \purple{the sensitivity-optimized rigging (SOR) method \cite{xu2014sensitivity}, and thereby significantly improving their results.}

Our approach is general and provides a universal scheme to add the shape dimension to any clothing pose model with a compact database. We validate our method with SOR~\cite{xu2014sensitivity} and a sequence of simulated clothing instances. Our results show that, with a small number of anchor points (approximately 150), our method can generate realistic clothing deformations with an extra time consumption of 4ms per frame. On a commodity CPU, we obtain $56$ FPS for a long-sleeved shirt with $12$K vertices. We also perform a preliminary user study in an immersive VR setting and highlight the perceptual benefits of our approach compared to prior interactive methods based on linear blend skinning. 

The rest of the paper is organized as follow: In Section \ref{sec:relatedworks}, we detail relevant related work in clothing animation. In Section \ref{sec:method}, we describe Taylor expansion, clothing shape model, clothing pose model, runtime synthesis, and database construction of our method. Section \ref{sec:experiment} details the results of our method. Discussion and conclusion will be given in Section \ref{sec:discussion} and Section \ref{sec:conclusion}.

\section{Related Work}\label{sec:relatedworks}

{\textbf{Physics-based clothing simulation.}}
In the past two decades, following the seminal work by Baraff et al. ~\cite{baraff1998large}, physics-based clothing simulation has become a hot topic in the computer graphics community. The following works focus on integration methods~\cite{hauth2003analysis,volino2005implicit,fierz2011element}, strain limiting ~\cite{provot1995deformation,goldenthal2007efficient,thomaszewski2009continuum,wang2010multi,ma2016anisotropic}, and various clothing simulation models \cite{grinspun2003discrete,english2008animating,choi2005stable,volino2009simple,bridson2002robust,harmon2008robust,zheng2012energy,cirio2014yarn,guo2018material}. While these methods can produce highly realistic  clothing deformations, they are typically quite time-consuming, especially with high-resolution cloth meshes. Many acceleration methods have been developed, including the projective dynamics method \cite{liu2013fast,bouaziz2014projective}, where integration is interpreted as an optimization problem; the Chebyshev semi-iterative approach, which accelerates the projective dynamics method \cite{wang2015chebyshev,fratarcangeli2018parallel}; and position-based dynamics~\cite{muller2007position}, where internal forces are replaced by position-based constraints to achieve both efficiency and stability.
\purple{Recently, parallel GPU-based methods have been developed for implicit integration and contact handling\cite{tang2016cama,tang2018cloth,tang2013gpu,govindaraju2005quick,pcloth}, which perform implicit integration and accurate collision handling (including self-collisions). The underlying collision queries are performed using CCD tests  that involve use of algebraic solvers and reliable computations~\cite{manocha1994algorithms,tang2014fast,manocha1998solving,manocha1992algebraic} for the elementary tests. The overall simulation algorithms exploit the parallelism on one or more GPUs and can accurately simulate at $2-10$ fps on high-end desktop GPUs. The actual running time can vary based on mesh resolution as well as the number of colliding configurations between triangles. Most VR applications require 30 fps (or higher performance) and current physics-based methods cannot offer such performance. Furthermore, in many applications, we need to perform cloth simulation on a mobile device (e.g., a smartphone) and we need methods that have a lower computational overhead.}
The time-consuming nature of collision processing has also prompted many researchers to focus on developing accelerated data structures such as bounding-volume hierarchies \cite{klosowski1998efficient}, distance fields \cite{fuhrmann2003distance}, shape approximation with simple primitives \cite{thiery2013sphere,wu2018variational}, and other spatial partitioning methods \cite{teschner2005collision}. Combined with position-based dynamics \cite{muller2007position}, these methods can achieve real-time clothing animation with plausible dynamics for a medium resolution mesh. However, the resulting clothing animation lacks detailed wrinkles.


{\textbf{Data-driven clothing animation. }}
Data-driven clothing animation techniques have received considerable attention in recent years for real-time applications that require high fidelity. De Aguitar et al. \cite{de2010stable} reduced the dimension of cloth space and body space with PCA and then learned a conditional dynamical model of cloth in the low-dimensional linear cloth space. Their method is fast and stable but cannot generate clothing deformations with highly detailed wrinkles. Wang et al. \cite{wang2010example} regarded wrinkles as a function of local joint angles and augmented coarse simulations with detailed wrinkles from a pre-computed wrinkle database. Kim et al. \cite{kim2013near} exhaustively searched a motion graph to generate realistic secondary motion of clothing deformations. However, both the memory space and the computational resources for the database are prohibitively high. Hahn et al. \cite{hahn2014subspace} simulated clothes in low-dimensional linear subspaces learned from performing PCA on clothing instances in different poses. While they can reproduce detailed folding patterns with only a few bases, the method is still too costly to be used in real-time clothing animation. To balance speed and quality, Xu et al. \cite{xu2014sensitivity} introduced real-time example-based clothing synthesis using sensitivity-optimized rigging to achieve physically plausible clothing deformation. Given a set of pre-computed example clothing deformations sampled at different poses, the method first rigs the example clothing shapes to their poses with the underlying body skeleton at each example pose in the offline stage. At runtime, the method synthesizes a clothing deformation of the input pose by blending skinned clothing deformations computed from nearby examples. \purple{Jin et al.\cite{jin2018pixel} represented clothing shapes as offsets from the underlying body surface, and performed convolutional neural networks to learn pose-dependent deformations in image space. Other methods directly used CNNs on triangle meshes, thereby predicting high resolution clothing deformations based on low resolution inputs~\cite{chentanez2020cloth}.}

In addition to the significant advances achieved in pose-dependent clothing animation, variability in the human body shape has also been considered in clothing animation for virtual try-on applications. Inspired by SCAPE \cite{anguelov2005scape}, Peng et al. \cite{guan2012drape} introduced a model of clothing animation called DRAPE (DRessing Any PErson), which separated clothing deformations due to body shape from those due to pose variation. DRAPE can fit avatars of various poses and shapes with customized garments and change the clothing model according to the body shape. \purple{Recently, many techniques have been proposed for learning-based clothing simulation. Wang et al.\cite{wang2018learning} learned a shared shape space in which users are able to indicate desired fold patterns simply by sketching, and the system generates corresponding draped garment and body shape parameters. However, their approach cannot generate clothing deformations corresponding to different poses.} Inspired by SMPL \cite{loper2015smpl}, Santesteban et al. \cite{santesteban2019learning} introduced a learning model of cloth drape and wrinkles. The key innovation of their method is that they added corrective displacements caused by body shape and pose variations to the template cloth mesh and then deformed the template mesh using a skinning function. In many ways, our approach is complimentary to this method. A limitation of their method is that the training dataset is prohibitively large, which is similar to other learning-based methods\cite{lahner2018deepwrinkles,patel2020tailornet}. \purple{By treating clothing as an extra offset layer from the body, Ma et al.\cite{ma2020learning} trained a conditional Mesh-VAE-GAN to learn the clothing deformation from the SMPL body model. This work has the limitation in that the level of geometric details they can achieve is upper-bounded by the mesh resolution of SMPL.} \purple{Yang et al.\cite{yang2018analyzing} modeled the clothing layer as an offset from the body and performed PCA to reduce the self-redundancies. In contrast, we perform PCA directly on the coordinates of clothing deformations to compute our clothing shape model.}
Many other works focused on cloth reconstruction from a single image or scan data\cite{saito2019pifu,natsume2019siclope,pons2017clothcap}, \purple{as well as cloth material recovery from video\cite{yang2017learning}}.

{\textbf{Sensitivity analysis. }}
Sensitivity analysis was originally used to determine the impact of the input data on the output results in linear programming problems \cite{saltelli2004sensitivity}, and it has been widely used to solve optimization problems in the field of graphics, including \purple{shell} design \cite{kiendl2014isogeometric}, composite silicone \purple{rubber} design \cite{zehnder2017metasilicone}, and robotics design \cite{geilinger2018skaterbots, zimmermann2019puppetmaster}. Sensitivity analysis was first introduced to the clothing simulation community by Umetani et al. \cite{umetani2011sensitive} to build up a mapping between variations of 2D patterns and 3D clothing deformations to achieve interactive clothing editing. After that, Xu et al. \cite{xu2014sensitivity} proposed a technique called sensitivity optimized rigging (SOR) to perform real-time clothing animation. They use sensitivity to both rig the clothing instances in example poses and find the example poses nearest to the input pose. In this paper, we use sensitivity to find the anchoring points nearest to the input pose and shape.

{\textbf{Taylor expansion. }}
A complex function can be approximated by its first order Taylor expansion in a neighborhood of the keypoint locations. Such a method has been applied in the field of simulation and animation to solve various approximation problems. To generate real-time facial animations, Barrielle et al. \cite{barrielle2019realtime} applied first-order Taylor approximation to the computations of the Singular Value Decomposition, thereby significantly accelerating the simulation of volumetric forces. Shen et al. \cite{shen2015geometrically} adopted a finite Taylor series approximation of the potential energy to avoid numerical singularity and instability during the simulation of inextensible ribbon. In the cloth simulation community, Taylor expansion is generally used to make the first order approximation of the internal force. To solve the nonlinear equation involved in the implicit backward Euler method, Baraff et al. \cite{baraff1998large} applied a Taylor series expansion to the force acting on the cloth and made the first order approximation, which leads to a linear system. As a result, their cloth simulation system can \purple{handle} large time steps in a stable manner. \purple{Chen et al.\cite{chen2010fully} proposed a fully geometric approach to simulate inextensible cloth that is subjected to a conservative force. They use Taylor expansion to linearize the constraint that preserves isometric deformations.} In this paper, we use Taylor expansion to factor the clothing deformation (which is a complex high-dimensional nonlinear function without an analytical expression) into two parts, as introduced by body shape and pose variations.

\section{Method}
\label{sec:method}

\subsection{Taylor Expansion For Clothing Deformation}
\label{subsec:method_overview}

\begin{figure*}[t]
\setlength{\abovecaptionskip}{0.0cm}
\centering
  \includegraphics[width=\textwidth]{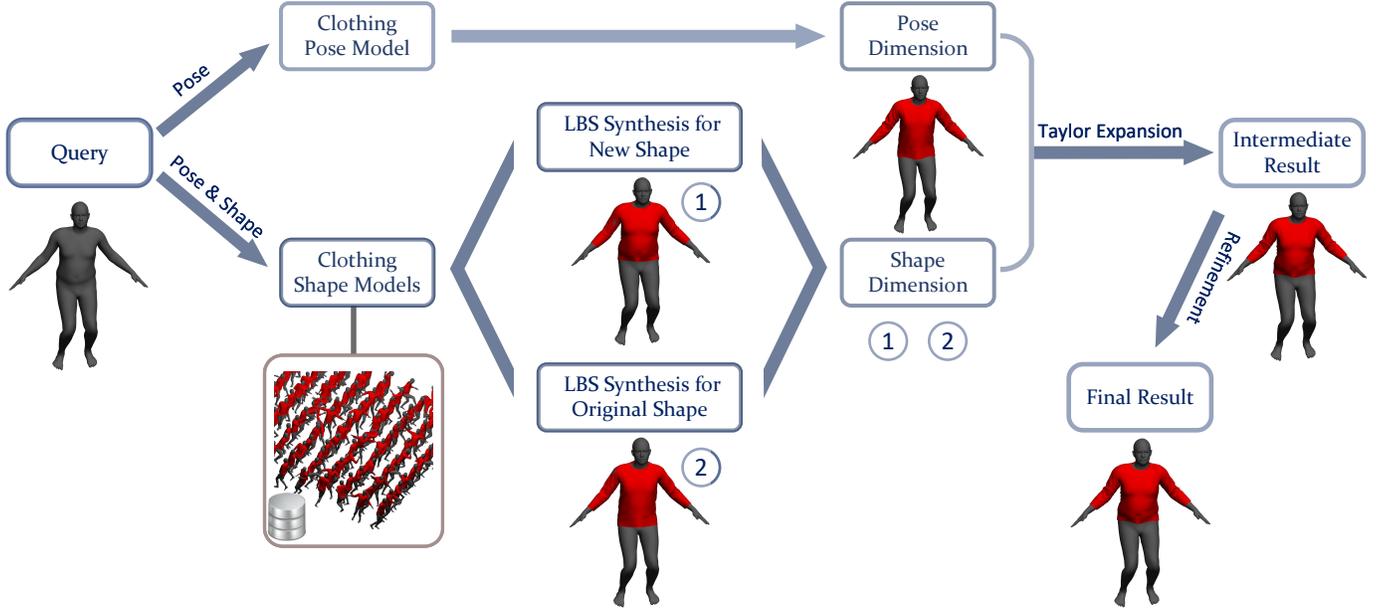}
  \caption{Overview of our clothing synthesis workflow. Given a query pose and shape, we first calculate the LBS synthesis results for the new shape and the original shape, and we refer to these as the shape dimension. The ``pose dimension'' is obtained from an external clothing pose model with the query pose and the original shape. Then we blend the ``shape dimension'' and the ``pose dimension'' using Taylor expansion to synthesize an intermediate result. Finally, we apply some refinement techniques such as penetration resolution to generate the final result.}
\label{fig:workflow}
\end{figure*}

Fig. \ref{fig:workflow} illustrates the pipeline of our approach. Our system consists of two stages: the offline training stage and the run-time clothing synthesis stage. To model the nonlinear deformation of clothing mesh under different poses and shapes, we precompute a pose training set of clothing meshes with a single template body in multiple poses and develop a shape training set fit to different body shapes in each pose. \purple{We use the pose training data to train a clothing pose model to predict the pose-dependent clothing deformation, given a template body (e.g. SOR); for each pose in the shape training data, we train a clothing shape model to predict the shape-dependent clothing deformation with a fixed pose}. At run-time, the method separates deformations induced by pose and shape (i.e. a clothing pose model and a clothing shape model) through Taylor expansion. Given a query body pose and shape, we first find nearby clothing pose models and clothing shape models and then blend the Taylor expansion result of each nearby anchoring point to synthesize an intermediate mesh. We then resolve penetrations and add damping to get the final clothing mesh for the input body.

Specifically, we represent the input body with shape parameters $\beta$ and pose parameters $\theta$. The clothing deformation $Y$ can be formulated as a high-dimensional function $Y = f(\beta, \theta)$ and approximated by the clothing on its nearby example body with shape parameters $\beta_0$ and pose parameters $\theta_0$ (we refer to $(\beta_0,\theta_0)$ as the anchoring point) using first-order Taylor expansion:
\begin{equation} \label{eq:taylorexpansionoriginal}
f(\beta, \theta)=f(\beta_0, \theta_0) + \Delta \beta f_\beta(\beta_0,\theta_0) + \Delta \theta f_\theta(\beta_0,\theta_0),
\end{equation}
where $\Delta \beta$ and $\Delta \theta$ are the shape and pose difference, respectively, between input body and its nearby example body. 

We use forward difference to calculate the partial derivatives $f_\beta(\beta_0,\theta_0)$ and $f_\theta(\beta_0,\theta_0)$ in Eq. \ref{eq:taylorexpansionoriginal} as follows:
\begin{equation}\label{eq:taylorexpansionderivative}
\begin{split}
f(\beta, \theta)=f(\beta_0, \theta_0) + \Delta \beta \cdot \frac{ f(\beta, \theta_0) - f(\beta_0, \theta_0)}{\Delta \beta}\ \\
+ \Delta \theta \cdot \frac{ f(\beta_0, \theta) - f(\beta_0, \theta_0)}{\Delta \theta}\ \\
= f(\beta_0,\theta) + (f(\beta,\theta_0)-f(\beta_0,\theta_0)),
\end{split}
\end{equation}
where $f(\beta_0,\theta)$ represents the clothing under a new pose with the body shape unchanged, which can be computed via a \textit{clothing pose model}. $f(\beta, \theta_0)$ denotes the clothing for a new body shape with the pose fixed, which can be calculated through a \textit{clothing shape model}. Correspondingly, $(f(\beta,\theta_0)-f(\beta_0,\theta_0))$ represents the shape-dependent cloth deformation, which is the cloth mesh deformation during the body shape change from $\beta_0$ to $\beta$ under the anchoring pose $\theta_0$.

In this way, the cloth deformations induced by body shape and pose can be separately computed and then combined. However, such an approximation cannot accurately obtain the clothing deformation under $(\beta, \theta)$ since the term of shape-dependent cloth deformation in Eq. \ref{eq:taylorexpansionderivative} should be measured under the new pose $\theta$ as $(f(\beta,\theta)-f(\beta_0,\theta))$, rather than under the anchoring pose $\theta_0$. To improve the approximation accuracy, we apply the Linear Blend Skinning (LBS) method to predict the cloth deformation for the new pose $\theta$ from the cloth mesh under its nearby sample pose $\theta_0$. Therefore, the shape-and-pose dependent cloth mesh deformation $f(\beta, \theta)$ can be formulated as the Augmented Taylor Expansion:
\begin{equation}\label{eq:taylorexpansionfinal}
f(\beta, \theta) = f(\beta_0,\theta) + \big(LBS^{\theta}_{\theta_0}(f(\beta,\theta_0))-LBS^{\theta}_{\theta_0}(f(\beta_0,\theta_0))\big),
\end{equation}
where $LBS^{\theta}_{\theta_0}$ stands for the linear blend skinning from pose $\theta_0$ to $\theta$.

Since Taylor expansion can only predict the function value at points that are not far away from the expansion point, we apply Taylor expansion at multiple points, and then blend these approximation results.

Note that both training and running a clothing pose model are time-consuming in practice. Taking SOR as an example, it takes about 30 hours to construct the necessary database and more than 10ms to predict the clothing deformation under a new pose, which is too time-consuming in our scenario since we need to run the clothing pose model once at each expansion point. To address this problem, we choose to apply Taylor expansion along the $\theta$ axis, i.e. expansion points have the same $\beta$ value, such as $(\beta_0,\theta_1)$, $(\beta_0,\theta_2)$, $(\beta_0,\theta_3)\cdots$, etc. In this way, for each expansion point $(\beta_0, \theta_s)$, the only thing we need to do is to train a clothing shape model at $\theta_s$, and we use only one clothing pose model $f(\beta_0,\theta)$ trained at $\beta_0$, referred to as the original shape. In our implementation, we set the shape parameters of the SMPL body model\cite{loper2015smpl} to be all zeros to get a medium stature body.

At run-time, given a query pose and shape, our method first finds the nearby clothing shape models through a sensitivity-based distance measure. For each nearby anchoring point, we compute the approximation result for the query pose and shape according to Eq. \ref{eq:taylorexpansionfinal}. After that, we blend the approximation results with blending weights computed from the distance measure.
In summary, our model predicts the clothing under any given body shape parameter $\beta$ and pose parameter $\theta$ by
\begin{equation}\label{eq:blending}
\begin{split}
f(\beta,\theta)=f(\beta_0, \theta) + \sum_{s=1}^{N_s}w^s LBS^{\theta}_{\theta_s}(f(\beta,\theta_s))\ \\
 - \sum_{s=1}^{N_s}w^s LBS^{\theta}_{\theta_s}(f(\beta_0,\theta_s)),
\end{split}
\end{equation}
where
$N_s$ is the number of clothing shape models and $w^s$ is the blending weight of the $s$-th data point. 
 Figure \ref{fig:workflow} illustrates the computation of this equation where the three items in the right-hand side \purple{of Eq. \ref{eq:blending}} are referred to as the \textit{pose dimension}, the \textit{LBS synthesis for new shape}, and the \textit{LBS synthesis for original shape}. 
 Details will be given in the following sections.

\subsection{Clothing Shape Model}\label{subsec:clothingshapemodel}
\begin{figure*}[t]
\setlength{\abovecaptionskip}{0.0cm}
\centering
  \includegraphics[width=\textwidth]{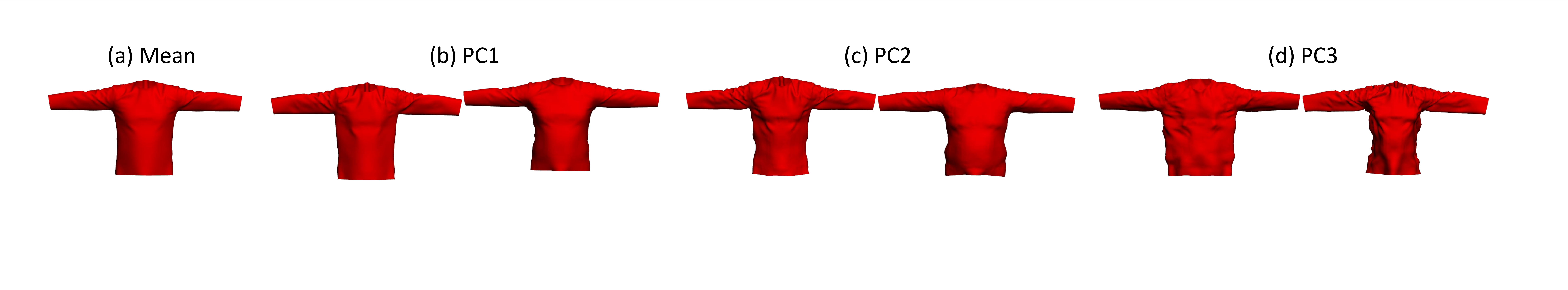}
  \caption{Clothing shape model. Deviations from the mean shape: (a) the average coordinates of the training data, referred to as the mean shape; (b-d) mean shape deformed along the first three principal component directions ( $\pm$ 3 standard deviations). Note that the global displacement in the vertical direction of (b) is introduced by the body height.}
\label{fig:principalcomponents}
\end{figure*}

The clothing shape model captures clothing deformation induced only by body shape. The model is learned from the clothing shape examples that are simulated under various body shapes with a fixed pose (each column in Figure~\ref{fig:clothingshapemodeltrainingdata}). Thus, the following procedure will be run independently for all anchoring poses, each to generate a clothing shape model for a specific pose. We use the SMPL parametric human model to represent the variations in human body shape. In the offline database construction, we select 17 training body shapes in the same pose. For each of the first 4 principal components of the body shape parameters $\beta$ in the SMPL model, we generate 4 body shapes ($\beta_k=-2,-1,1,2$), keeping the remaining parameters in $\beta$ as 0. To these 16 body shapes, we add the nominal shape with $\beta=0$.



For each generated body shape, we perform a one-second simulation to completely drape the clothing on the avatar. All simulations are produced using a clothing model that combines the StVK membrane model\cite{volino2009simple} and the isometric bending model\cite{bergou2007tracks} to simulate a garment example. It is worth noting that in the clothing simulation, the patterns and characteristics of the garment mesh do not change for all body shapes.

For the $k$-th generated clothing instances, we concatenate the clothing coordinates of all vertices into a single column vector $\vec{c^k}$. Then all 17 clothing samples are collected into a matrix $S=[\vec{c^1},...,\vec{c^k},...,\vec{c^{17}}]$. Principal component analysis (PCA) is used to find a low-dimensional subspace so that $\vec{c^k}$ can be approximated by the following equation:
\begin{equation} \label{eq:phipca}
\vec{c^k}=U\vec{\phi}^k+\vec{u},
\end{equation}
where $\vec{u}$ is the mean coordinates of clothing meshes and $\vec{\phi}^k$ is the clothing shape coefficients to represent the clothing shape $\vec{c^k}$. The matrix $U$ represents the first few principal components of the shape deformation space. As shown in Figure \ref{fig:pcaconvergence}, the variance converges around 5 principal components, which is 98.6\% of the total variance. Moreover, we find that the remaining principal components have little effect on the final result. Therefore, we use the first 5 principal components in our clothing shape model to balance efficiency and accuracy. Figure \ref{fig:principalcomponents} illustrates the mean and first three principal components for a men's long-sleeved shirt.

Given 17 body and clothing training pairs, we learn a linear mapping, $W$, between body shape parameters and clothing shape parameters using L2-regularized least squares with the weight of the regularized term being 0.2.

For an input body shape $\beta$, the corresponding clothing shape coefficients $\vec{\phi}$ can be predicted by:
\begin{equation} \label{eq:beta2phi}
 \vec{\phi} = W\cdot \left(\vec{\beta}, \vec{\beta}^2, 1 \right)^T.
\end{equation}

Thereafter, the clothing shape deformation under the body shape $\beta$ can be predicted through $\vec{\phi}$ using Eq. \ref{eq:phipca}. In practice, we find that there might be some minor interpenetrations between the predicted clothing mesh and the body mesh. We resolve this by pushing the implicated clothing vertex \purple{in the direction of} the normal of its closest body vertex until the interpenetration is settled. This should not be confused with the penetration handling step in Sec. \ref{subsubsec:penetrationhandling}.

\subsection{Clothing Pose Model}\label{subsec:clothingposemodel}

The clothing pose model captures clothing deformation introduced only by pose changes. Many excellent clothing pose models have been presented over the last decade \cite{kim2013near,hahn2014subspace,xu2014sensitivity}. Our animation scheme
does not limit the type of clothing pose model; in extreme cases, we apply a simulated sequence as the clothing pose model (see Sec. \ref{subsec:animationresult}). For real-time virtual try-on applications, we adopt the sensitivity-optimized rigging method proposed in \cite{xu2014sensitivity} since it has a good balance of accuracy and speed. 

The system consists of two phases: the offline rigging stage and the run-time clothing synthesis stage. Given a set of pre-computed example clothing deformations sampled at different poses, we first rig the example clothing shapes to their poses with the underlying body skeleton bones at each example pose in the offline stage. At run-time, a clothing deformation of the input pose is synthesized by blending skinned clothing deformations computed from its nearby examples.

\begin{figure*}[t]
\setlength{\abovecaptionskip}{-0.0cm}
\centering
  \includegraphics[width=\textwidth]{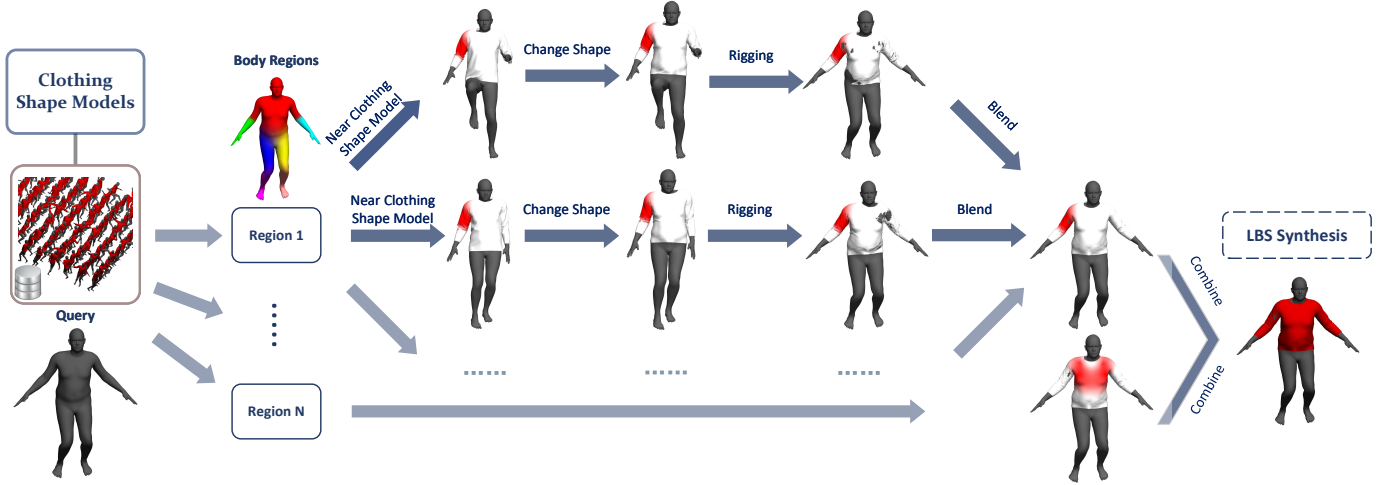}
  \caption{The LBS synthesizing process. Given a query pose and shape, we first choose nearby clothing shape models. For each chosen clothing shape model, we change its shape to the query shape (the original shape), and deform the clothing mesh to the query pose using a linear blend skinning. Finally, we blend these deformations to get what we call the \textit{LBS synthesis for the query shape (the original shape).}}
\label{fig:lbssynthesis}
\end{figure*}

\subsection{Runtime Synthesis}\label{subsec:runtimeshythesis}

Given an input body with shape parameters $\beta$ and pose parameters $\theta$, our method first finds nearby example poses through a sensitivity-based distance measure and then approximates the clothing deformation using Taylor expansion, as shown in Eq. \ref{eq:blending}.  Specifically, the first term $f(\beta_0,\theta)$ computes the clothing under the input pose with the original body shape, which can be predicted through the clothing pose model described in Sec. \ref{subsec:clothingposemodel}.

The computation of the second term $\sum_{s=1}^{N_s}w^s LBS^{\theta}_{\theta_s}(f(\beta,\theta_s))$ consists of three steps: predicting new clothing instances $f(\beta,\theta_s)$ for each clothing shape model, applying LBS to $f(\beta,\theta_s$), and blending the LBS results.

For each clothing shape model, we first calculate the clothing shape coefficients through Eq. \ref{eq:beta2phi} using the input shape $\beta$. Then we compute the new clothing instances $f(\beta,\theta_s)$ by Eq. \ref{eq:phipca}.

To apply LBS to new clothing instances $f(\beta,\theta_s)$, we first find the closest body vertex for each clothing vertex and set the bone weights, $w_b^s$, of a clothing vertex to that of its closest body vertex. We refer to this step as the \textit{binding information updating step}. Then we deform each nearby clothing mesh $f(\beta, \theta_s)$ towards the query pose as $\bar{y}^s = \sum_{b=1}^{N_b} w_b^s \big (R_b^{\theta_s,\theta}y^s+T_b^{\theta_s,\theta} \big )$,
where $R_b^{\theta_s,\theta}$ and $T_b^{\theta_s,\theta}$ are the relative rotation and translation of bone $b$ from example pose $\theta_s$ to input pose $\theta$, respectively, and $w_b^s$ is the bone weight defined on the clothing vertex $y^s$ of $f(\beta, \theta_s)$. We denote this equation as $LBS^{\theta}_{\theta_s}(f(\beta,\theta_s))$, as shown in Eq. \ref{eq:blending}.

We use a sensitivity-based distance measurement both to find nearby clothing shape models and to compute the blending weights $w^s$. To reduce the required clothing shape models in the database, we divide the clothing mesh into several regions, as in \cite{xu2014sensitivity}. To this end, we manually partition the bones into $N_g=7$ regions (shown on the top left of Figure \ref{fig:lbssynthesis}). A region weight $w_{g,y}$ for a clothing vertex $y$ is computed by summing the bone weights $w_b^0$ for the bones of the current region. $w_b^0$ is computed in the T-pose $s=0$ for $\beta$ through the \textit{binding information updating step}. In this way, our method synthesizes the result for each region separately.

For each region $g$, we compute the sensitivity-based distance $D_g^s(\theta)$ between the input pose $\theta$ and the pose of the $s$-th data point as the weighted sum of differences of joint angles:
\begin{equation}\label{eq:weighteddiffonejoint}
\begin{split}
D_g^s(\theta) = \sum_{y\in Y} w_{g,y} \sum_{m=1}^{3N_L} ||\bar{s}_{y,m}\cdot \Theta_m(\theta,\theta_s)||^2 \\
=\sum_{m=1}^{3N_L} Q_{g,m} ||\Theta_m(\theta,\theta_s)||^2,
\end{split}
\end{equation}
where superscript $s$ indicates values for the $s$-th clothing shape model, $N_L$ is the number of joints, and $\Theta_m(\theta,\theta_s)$ calculates the $m$-th joint angle difference. $\bar{s}_{y,m} = \frac{1}{17} \sum_{k=1}^{17} ||s_{y,m}^k||$ is the average sensitivity of 17 training shapes, where $s_{y,m}^k$ indicates the three-dimensional coordinate differences of a clothing vertex $y$ under a small joint rotation of $m$-th joint angle, calculated under T-pose and the $k$-th training shape. $Q_{g,m}=\sum_{y\in Y} w_{g,y}||\bar{s}_{y,m}||^2$ reflects the influence of the $m$-th joint angle on region $g$. $\bar{s}_{y,m}$ indicates the influence of the $m$-th joint angle on clothing vertex $y$, which is computed in the database construction stage. Each time we change the body shape, we update $Q_{g,m}$ for each region once according to the new region weights of clothing vertices. At run-time, we can compute the sensitivity-based distance $D_g^s(\theta)$ efficiently since we only have to calculate the joint angle differences $\Theta_m(\theta,\theta_s)$.

Given the distance $D_g^s(\theta)$, the weight for the region is calculated as $W_g^s(\theta)=1/(D_g^s(\theta) + \epsilon)^k$, where $\epsilon$ is a small number in case of zero division and $k$ regulates the influence of closer examples. A small $k$ tends to smooth the animation and lose fine wrinkles, while a large $k$ tends to preserve fine details but results in discontinuity. In our implementation, we set $k=3$. In practice, we use the first five nearby clothing shape models to synthesize the final result, i.e. we set $W_g^s(\theta)=0$ except for the top five largest ones.


Finally, the blending weight for each clothing vertex in example $s$ is calculated as:
\begin{equation}\label{eq:vertexblendingweight}
w^s = \sum_{g=1}^{N_G}\big( w_g W_g^s(\theta)/\sum_{s=1}^{N_s}W_g^s(\theta) \big).
\end{equation}

 The computation pipeline of the third term $\sum_{s=1}^{N_s}w^s LBS^{\theta}_{\theta_s}(f(\beta_0,\theta_s))$ is basically the same as the second term except for the body shape. In other words, we first compute clothing instances under the original shape $\beta_0$ for each clothing shape \purple{model}, denoted as $f(\beta_0,\theta_s)$. Then we blend the LBS results of $f(\beta_0, \theta_s)$ using the same weights as in the second term. Note that $f(\beta_0,\theta_s)$ and their binding information are computed only once for an application.

\subsection{Refinement}
\subsubsection{Decaying Effects}
\begin{figure}[t]
\setlength{\abovecaptionskip}{0.0cm}
\centering
  \includegraphics[width=0.48\textwidth]{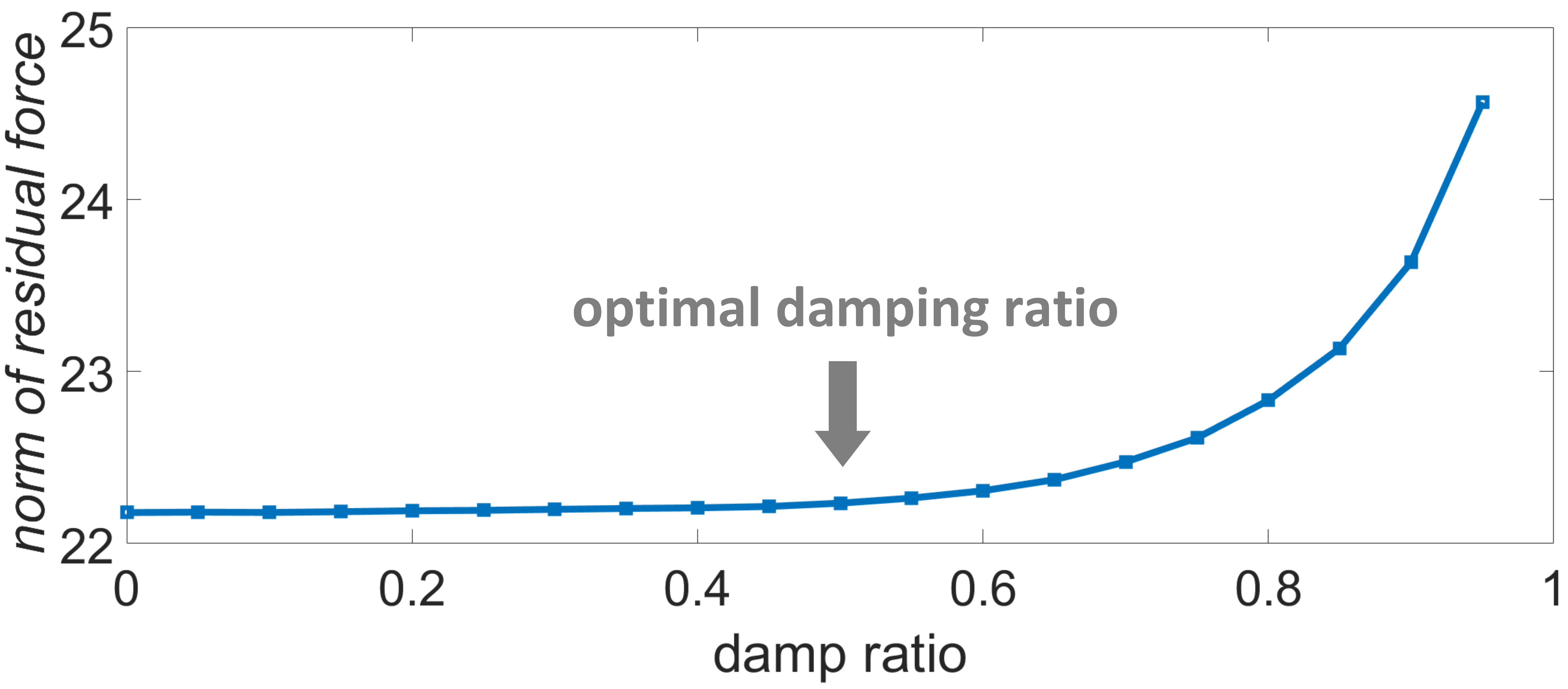}
  \caption{Relationship between the damping ratio and the norm of the residual. We choose the maximum damping ratio that keeps the norm of the residual low.}
  \label{fig:choosedampingratio}
\end{figure}

In practice, we find our clothing synthesis result may experience sudden changes for sudden input pose changes. Similar to \cite{xu2014sensitivity}, we prevent such a problem by blending the distance at the current time step $D_{g}^{s}$ with that of the previous time step $D_g^{\prime s}$:
\begin{equation}\label{eq:decaying}
D_g^s = \eta D_g^{\prime s} + (1-\eta)D_g^s,
\end{equation}
where $\eta$ is the damping ratio, ranging from 0 to 1. To determine the damping ratio, we first calculate the relationship between the damping ratio and the norm of residual, computed under the original body shape. Then we choose the maximum value of the damping ratio while the norm of the residual is still low (see Figure \ref{fig:choosedampingratio}). The rationale is that the higher the value of the damping ratio, the stronger the decaying effect and the less flickering.
We believe this simple technique will introduce some hysteresis while maintaining a natural result.

\subsubsection{Penetration Handling}\label{subsubsec:penetrationhandling}

\begin{figure}[t]
\setlength{\abovecaptionskip}{0.0cm}
\centering
  \includegraphics[width=0.48\textwidth]{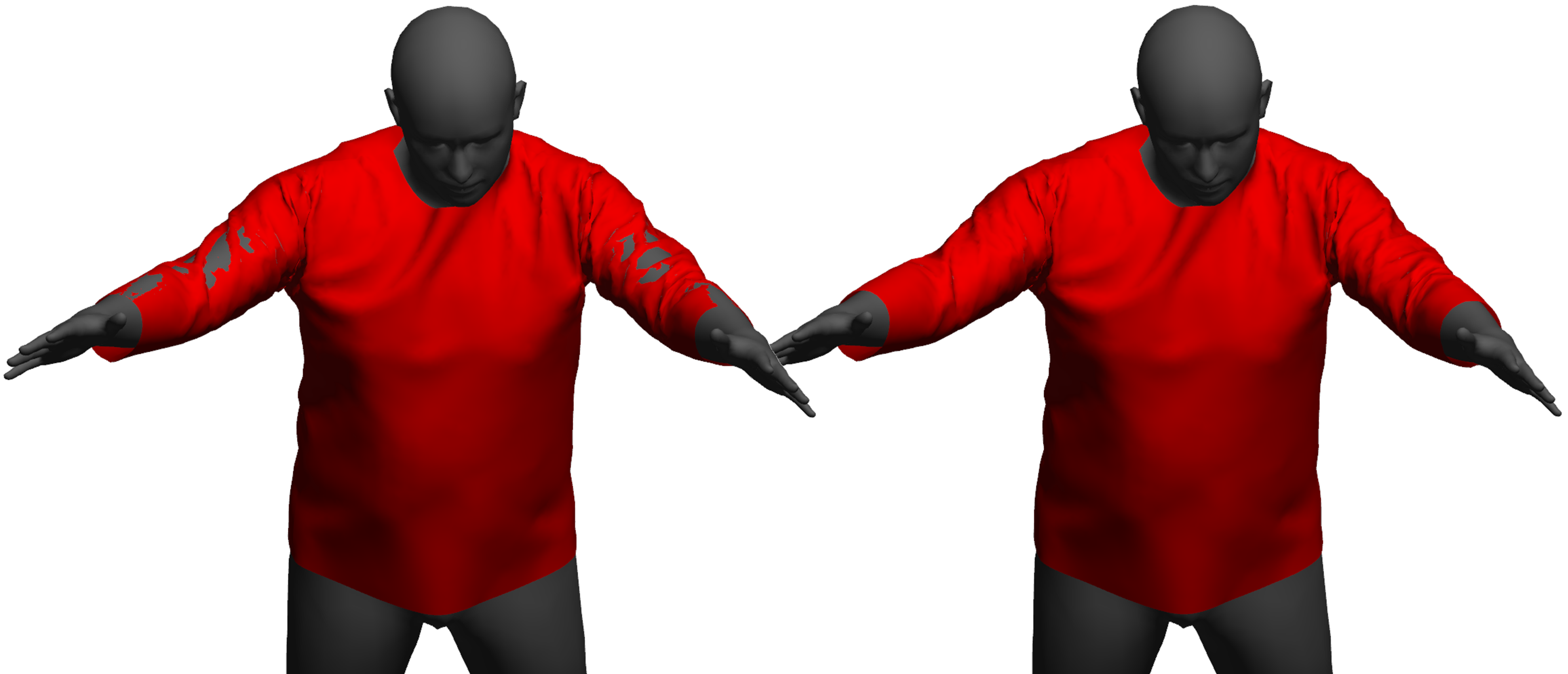}
  \caption{The result before (left) and after (right) our penetration handling process.}
  \label{fig:penetrationhandling}
\end{figure}

After we get the synthesized result through Eq. \ref{eq:blending}, there might be interpenetrations between the clothing mesh and the body mesh (as shown in Figure \ref{fig:penetrationhandling}). Like SOR \cite{xu2014sensitivity}, we use a re-projection technique with three steps to resolve this problem.

First, every time we change the shape of the avatar, for each clothing shape model we re-calculate the initial \purple{distance of a clothing vertex from} its closest body vertex. We refer to this as the initial clearance.
Second, in the run-time stage, if the clearance between a clothing vertex and its closest body vertex is less than the initial clearance, for each nearby clothing shape model, we re-project the clothing vertex towards the direction of the normal of its closest body vertex as:
\begin{equation}\label{eq:penetrationhandling}
\begin{cases}
\hat{y}^s = y + d \cdot \vec{n} \\
d^s = max (0,\quad d_0^s - h^s),
\end{cases}%
\end{equation}
where $y$ is a clothing vertex synthesized by Eq. \ref{eq:blending}, $\vec{n}$ is the normal of the closest body vertex of $y$, and $h^s$ is current clearance. We set $d_0^s=min\{h_0^s,\epsilon_p\}$, where $h_0^s$ is the initial clearance, and $\epsilon_p$ is to mimic the penetration depth margin in cloth simulation (in our implementation, we empirically set $\epsilon_p = 5mm$).
Finally, we blend the re-projection results of all nearby examples as $\bar{y}=\sum_{s=1}^{N_s}w^s\hat{y}^s$.

\subsection{Database Construction}
\label{subsec:database}

\begin{figure}[t]
\setlength{\abovecaptionskip}{0.0cm}
\centering
  \includegraphics[width=0.48\textwidth]{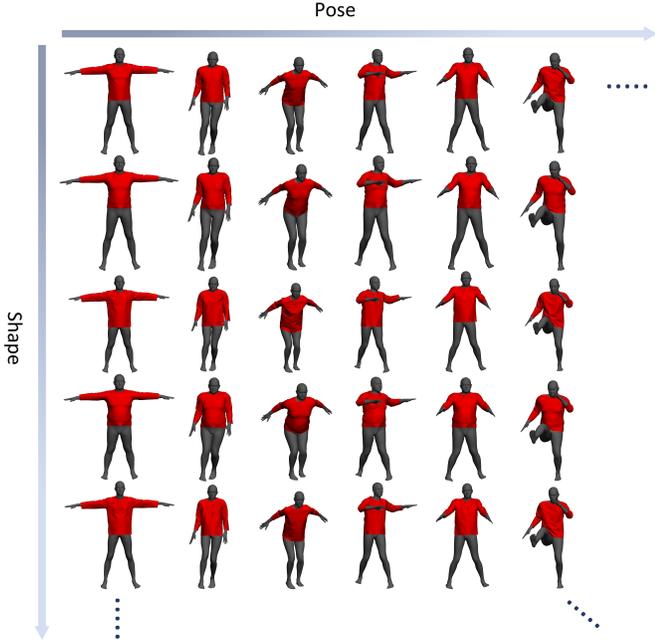}
  \caption{Part of our training data. Each column is the training data for a clothing shape model.}
  \label{fig:clothingshapemodeltrainingdata}
\end{figure}

To generate our example database, we first select 32 motion sequences from the CMU motion capture library \cite{cmumotiondata} and sample a pose for every four frames. In total, we obtain 17k different poses representing the whole pose space. Then we use weighted K-means to classify these poses into a certain number of clusters, which will be used to generate our example database of the clothing shape model.  In our implementation, it takes about one hour to classify these poses into a typical value of 150 clusters. 

The weight of a joint should reflect its importance in clothing animation. For instance, while the rotation of the knee joint has little, if any, influence on the animation result of a T-shirt, it plays a crucial role in the deformation of pants. To this end, we use the sum of the norm of the sensitivity of a joint, $s^L$, as its weight in the clustering process, calculated as
\begin{equation}\label{eq:kmeansweight}
s^L = \sum_{y\in Y}\sum_{m=1}^{3}||\bar{s}_{y,m}||,
\end{equation}
where $m$ represents the degree of freedom of joint $L$ and $\bar{s}_{y,m}$ is the same as in Eq. \ref{eq:weighteddiffonejoint} (i.e. $s^L$ is computed over all the training shapes).

The poses in each row of Figure \ref{fig:clothingshapemodeltrainingdata} are the clustering result of a male wearing a long-sleeved shirt. The cluster centers are essentially the anchoring points for Taylor expansion in Eq. \ref{eq:blending}. For each anchoring point, we generate a clothing shape model, which we elaborate on in Sec. \ref{subsec:clothingshapemodel}.

\section{Experiments}
\label{sec:experiment}
We implemented our approach in C++ and reported its performance on an off-the-shelf computer with an Intel Core i7-7700K CPU 4.20GHz and 16GB memory. The same material parameters are applied for all types of clothing: bending stiffness is $10^{-5} N/m$  , stretching stiffness is $30 N/m$, area density is $0.1 kg/m^2$, and dynamics and static coefficient of friction is $0.3$.
\subsection{Database Construction and Run-time Performance}
\label{subsec:databaseconstruction}
\begin{table}[htbp!]
\begin{center}
\resizebox{0.5\textwidth}{!}
{
\begin{tabular}{|c|c|c|c|}
\hline
Clothing & Long-sleeved shirt &  T-shirt & Pants  \\
\hline
number of vertices & 12.2k & 12.0k & 11.7k \\
number of triangles & 23.4k & 23.8k & 22.4k \\
\hline
number of data points & 150 / 150 & 150 / 150 & 150 / 170 \\
database size (MB) & 208.5 / 56.7  & 206.3 / 56.0 & 199.5 / 61.7 \\
\hline
construction time (hrs) & 15 / 7 & 10 / 6  & 14 / 8\\
runtime frame rate (FPS) & 56 & 55 & 71 \\
\hline
\end{tabular}
}
\end{center}
\caption{Statistics for three different clothing databases. The number to the left of ``/'' is the data for our method while the number to the right of ``/'' indicates the data for the external clothing pose model we use, i.e. SOR. Note that we have discarded the translation item in SOR and replaced the sampling method with ours. These measures contribute to the high speed of the database generation process of SOR. }
\label{tbl:statistics}
\end{table}

We create databases for three clothing models: a long-sleeved shirt and pants for a male body and a T-shirt for a female body (see Figures \ref{fig:teaser} and \ref{fig:moreresult}).
For each clothing model, we first calculate its joint weights using Eq. \ref{eq:kmeansweight}; then we use weighted K-means to obtain anchoring points; finally, we generate a clothing shape model for each anchoring point. Based on the independence of the anchoring points, our data points can be generated in parallel. \purple{Take T-shirt as an example, we generate $150 \times 17 + 150$ (the data used for our method + the data for SOR) clothing instances to generate our database. On the other hand, \cite{santesteban2019learning} simulated $7117 \times 17$ clothing instances. That is, the size of their database is more than 40 times larger than ours.}
Table \ref{tbl:statistics} shows the details of databases constructed for these clothing models.

\subsection{Separability}
\label{subsec:separability}
\begin{figure*}[t]
\centering
  \includegraphics[width=\textwidth]{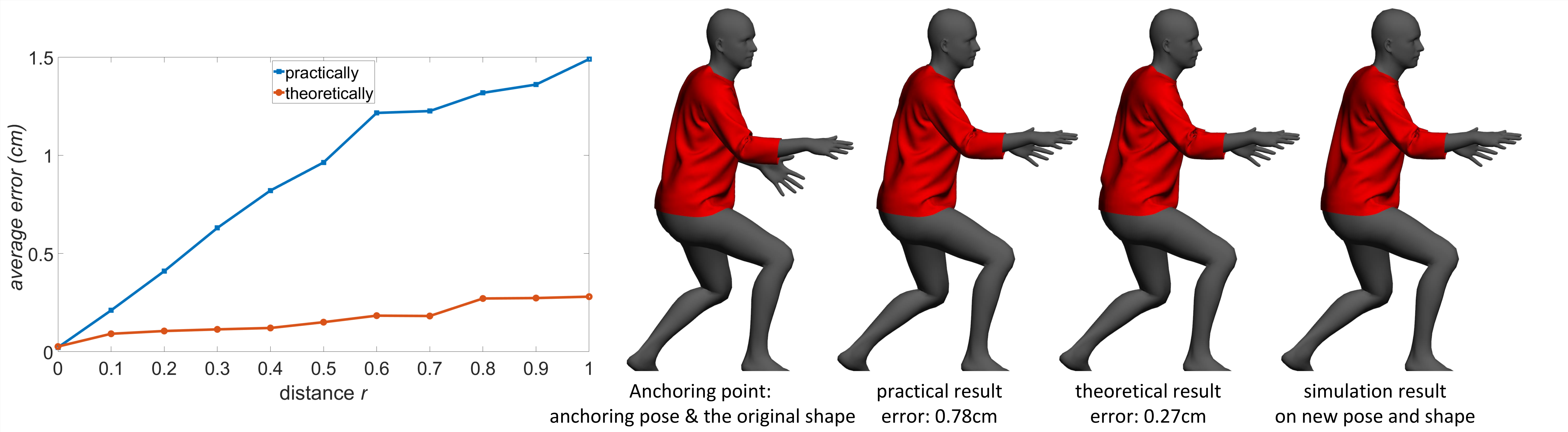}
  \caption{Validation of the local linear separability of pose-dependent deformation and shape-dependent deformation. Practically, we use Eq. \ref{eq:taylorexpansionfinal} to compute the synthesized clothing and, theoretically, we replace each term on the right-hand side of Eq. \ref{eq:taylorexpansionfinal} with the corresponding simulation result. As shown in the graph, both approximation errors converge to zero when the distance to the anchoring point approaches zero. The images on the right are the results for a randomly generated pose and shape with $r=0.5$. Both results (the middle two) can realistically recover the folds and wrinkles (the right-most figure is the ground truth) while the theoretical result has a smaller error.}
  \label{fig:separability}
\end{figure*}
In this section, we validate the local linear separability of pose-dependent deformation and shape-dependent deformation. For an anchoring point, we uniformly sample $m$ neighboring data points (we set $m = 100$ in our experiment), all of which are a distance $r$ from the anchoring point. Specifically, we randomly generate $3N_L+4$ (number of joint angles plus 4 shape parameters) numbers that are then concatenated into a vector $\vec{e}$. We normalize $\vec{e}$ in terms of 2-norm and then multiply it by $r$. The first $3N_L$ components of the final $\vec{e}$ are the displacement of joint angles from the anchoring point while the last $4$ components are the displacement of the shape parameters. Then, for each neighboring point, we calculate the average Euclidean vertex distance between the simulation result and the synthesized result as the approximation error. The average approximation error of all neighboring points is regarded as the approximation error of the anchoring point at distance $r$.

We run this procedure both practically and theoretically. Practically, we apply Eq. \ref{eq:taylorexpansionfinal} to compute the synthesized clothing; theoretically, we replace each term on the right-hand side of Eq. \ref{eq:taylorexpansionfinal} (i.e. $f(\beta_0,\theta); f(\beta,\theta_0)$; and $f(\beta_0,\theta_0)$) with the corresponding simulation result. Hence, in the practical experiment, the error comes from three sources: the clothing pose model, the clothing shape model, and the linear approximation. In the theoretical experiment, the error is caused only by the linear approximation.  As shown in Figure \ref{fig:separability}, for a randomly chosen anchoring point, both results can recover the folds and wrinkles with good fidelity; the error of the theoretical experiment is smaller than that of the practical experiment given the same distance $r$. Meanwhile, when $r$ is small, both approximation errors are relatively small. The errors converge to zero when $r$ approaches zero. This demonstrates the approximate separability of pose-dependent deformation and shape-dependent deformation.

\subsection{Improvement on SOR by applying our sampling method}
\label{subsec:improvementonsor}
\begin{figure}[t]
\setlength{\abovecaptionskip}{0.0cm}
\centering
  \includegraphics[width=0.48\textwidth]{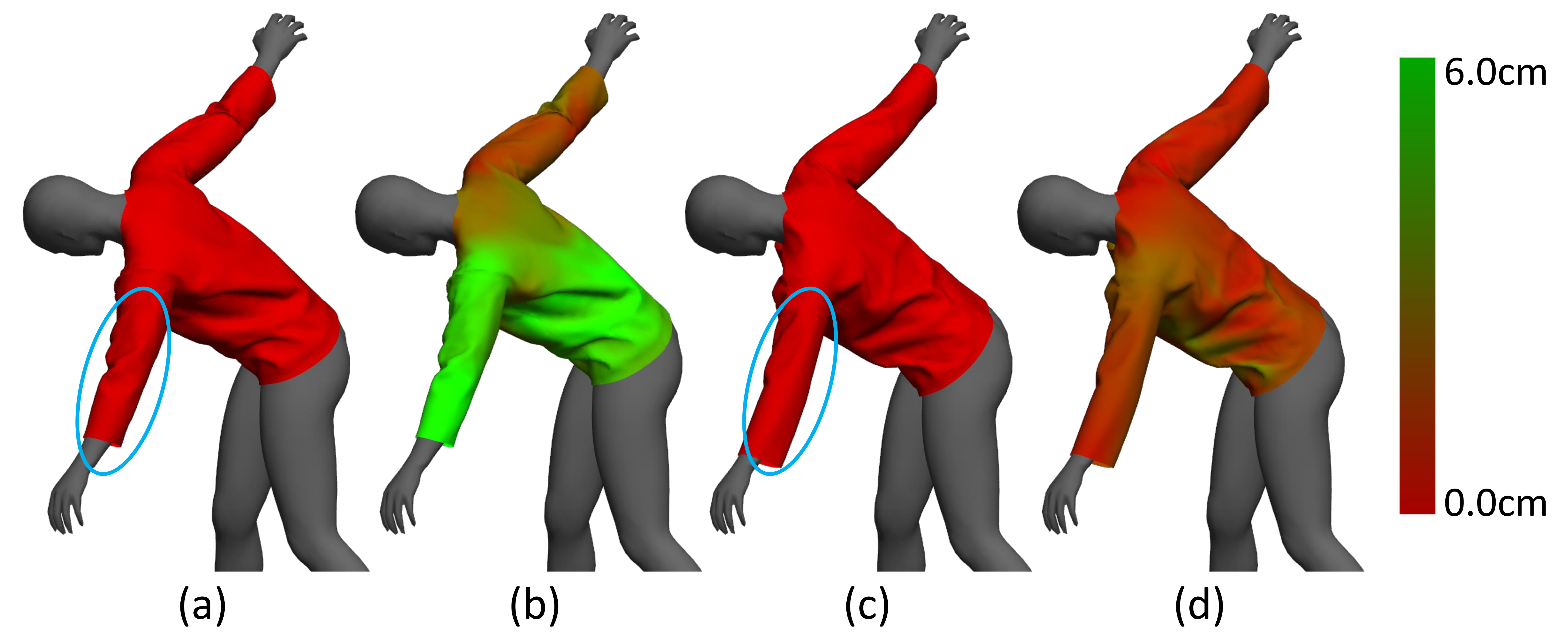}
  \caption{Qualitative comparison between the original SOR and our method: (a) original SOR result; (b) error between original SOR result and the simulation result; (c) our result; (d) error between our result and the simulation result. The errors are marked in green. We can see that our result is closer to the ground truth and looks more natural, especially in the sleeve area.}
  \label{fig:sorimprovementvisual}
\end{figure}

\begin{table*}[htbp!]
\begin{center}
\begin{tabular}{|l|c|c|c|c|c|c|c|c|c|c|c|c|c|c|c|c|c|c|}
\hline
$n_{dp}$ & 1 & 10  & 20  & 30  & 40  & 50  & 60 & 70 & 80 & 90 & 100 & 110 & 120 & 130 & 140 & 150\\
\hline
SOR & 45.6 & 28.0 & 25.7 & 28.1 & 29.0 & 29.8 & 28.9 & 26.0 & 25.8 & 25.2 & 23.7 & 23.3 & 23.1 & 22.6 & 22.3 & 22.2\\
\hline
Ours & 27.6 & 20.5 & 20.0 & 18.4 & 16.5 & 18.7 & 18.2 & 17.3 & 17.4 & 18.9 & 17.6 & 17.2 & 16.4 & 16.9 & 17.4 & 17.2\\
\hline
Imp & 39\% & 27\% & 22\% & 34\% & 43\% & 37\% & 37\% & 33\% & 33\% &    25\% & 26\% & 26\% & 29\% & 25\% & 22\% & 22\%\\
\hline
\end{tabular}
\end{center}
\caption{Quantitative comparison between the original SOR and our method. $n_{dp}$ stands for the number of data points and ``Imp'' is an abbreviation of ``improvement''. We can see that our sample scheme can reduce the norm of the residual by more than 22\%, meaning that the synthesized clothing deformations of our method are closer to the equilibrium states.}
\label{tbl:sorimprovementresidual}
\end{table*}


Each step of MCMC sampling in SOR aims to find the pose that maximizes the norm of the residuals of the synthesized clothing, i.e. they try to minimize the maximum error. However, a pose with the biggest error is sometime not a natural human pose. On the other hand, our sampling method can generate poses that are representative of real life. Therefore, we believe our sampling method is better than the one used in SOR. To demonstrate this, we replace the sampling scheme in SOR with our sampling method and regenerate the database for SOR. The weights of joints in the clustering process are computed with the sensitivity under the original body shape from SOR. Qualitatively, our method can generate more natural clothing deformation, as shown in Figure \ref{fig:sorimprovementvisual}. Quantitatively, as shown in Table \ref{tbl:sorimprovementresidual}, our method reduces the norm of the residual force by over 22\%, as computed over the 32 motion sequences described in Sec. \ref{subsec:databaseconstruction}. Furthermore, the sampling points generated by our sampling method are independent of each other, which enables the database to be generated in parallel.


\subsection{Clothing Shape Model}


\begin{figure}[t]
\setlength{\abovecaptionskip}{0.0cm}
\centering
  \includegraphics[width=0.48\textwidth]{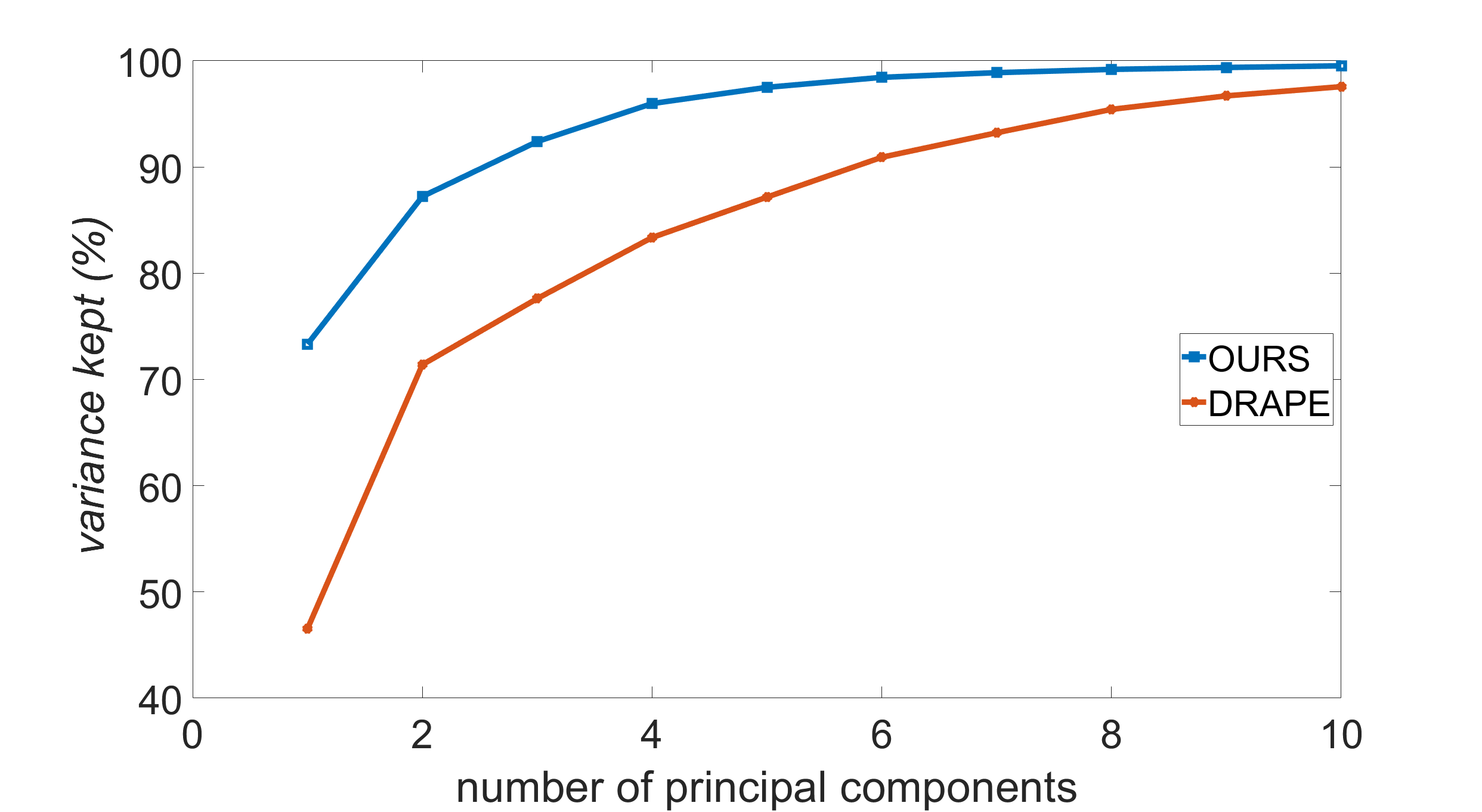}
  \caption{Convergence of variance during PCA of the training data (left most column in Figure \ref{fig:clothingshapemodeltrainingdata}). Our method applies PCA to coordinates of clothing vertices while DRAPE applies PCA to deformation gradients of clothing triangles. We can see that our method converges faster and captures more variance with a given number of principal components.}
\label{fig:pcaconvergence}
\end{figure}



We compare our clothing shape model to that of DRAPE \cite{guan2012drape}. Instead of using the SCAPE body \cite{anguelov2005scape}, we map the first four shape parameters of SMPL \cite{loper2015smpl} to their clothing shape parameters and use our training data to train this mapping function. We take a men's long-sleeved shirt as the representative clothing type for quantitative experiments of clothing shape models. The results are similar for other clothing types. To better evaluate the performance of the clothing shape model, we use the average Euclidean vertex distance to measure the prediction error and compute the average prediction error for 100 randomly generated body shapes.

\begin{figure*}[t]
\setlength{\abovecaptionskip}{0.0cm}
\centering
  \includegraphics[width=0.98\textwidth]{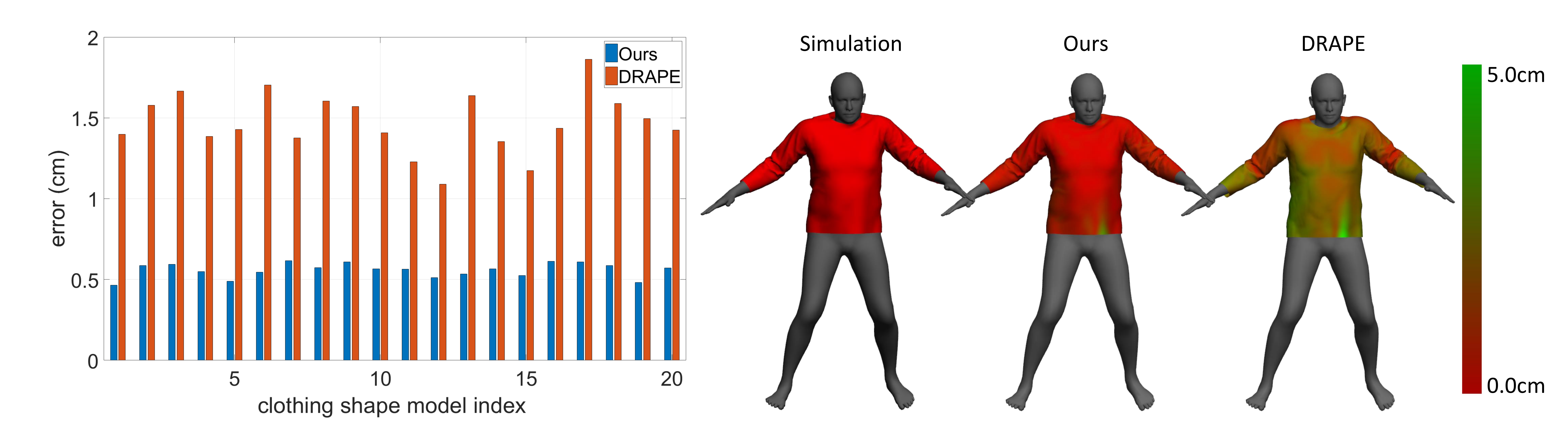}
  \caption{Comparison of our clothing shape model and that of DRAPE. The bar graph on the left shows the errors for 20 clothing shape models. The prediction results for the second clothing shape model for a random body shape are shown on the right. The errors are marked in green. }
\label{fig:clothingshapemodelscomparison}
\end{figure*}

Figure \ref{fig:clothingshapemodelscomparison} (left) illustrates the average errors for 20 different clothing shape models (20 different poses). Compared to DRAPE, our method reduces the error for each clothing shape model by 60\%. This can also be seen in the prediction results of the second clothing shape model for a random body shape; please see the right part of Figure \ref{fig:clothingshapemodelscomparison}. As shown in Figure \ref{fig:pcaconvergence}, our method captures more variance than DRAPE given the same number of principal components (98.6\% vs. 87.2\% when using the first 5 principal components), which contributes to the higher performance of our clothing shape model. It is also worth noting that the square term $\beta^2$ reduces the prediction error by 3\%.

Furthermore, our mapping function from body shape to clothing deformation only involves multiplication and addition, and thus is more efficient to calculate than DRAPE, where an optimization problem needs to be solved. In practice, it takes 22ms for our clothing shape model to predict the new clothing information, compared to 300ms for DRAPE.


\subsection{Animation Results}\label{subsec:animationresult}

\begin{figure*}[t]
\setlength{\abovecaptionskip}{0.0cm}
\centering
  \includegraphics[width=0.92\textwidth]{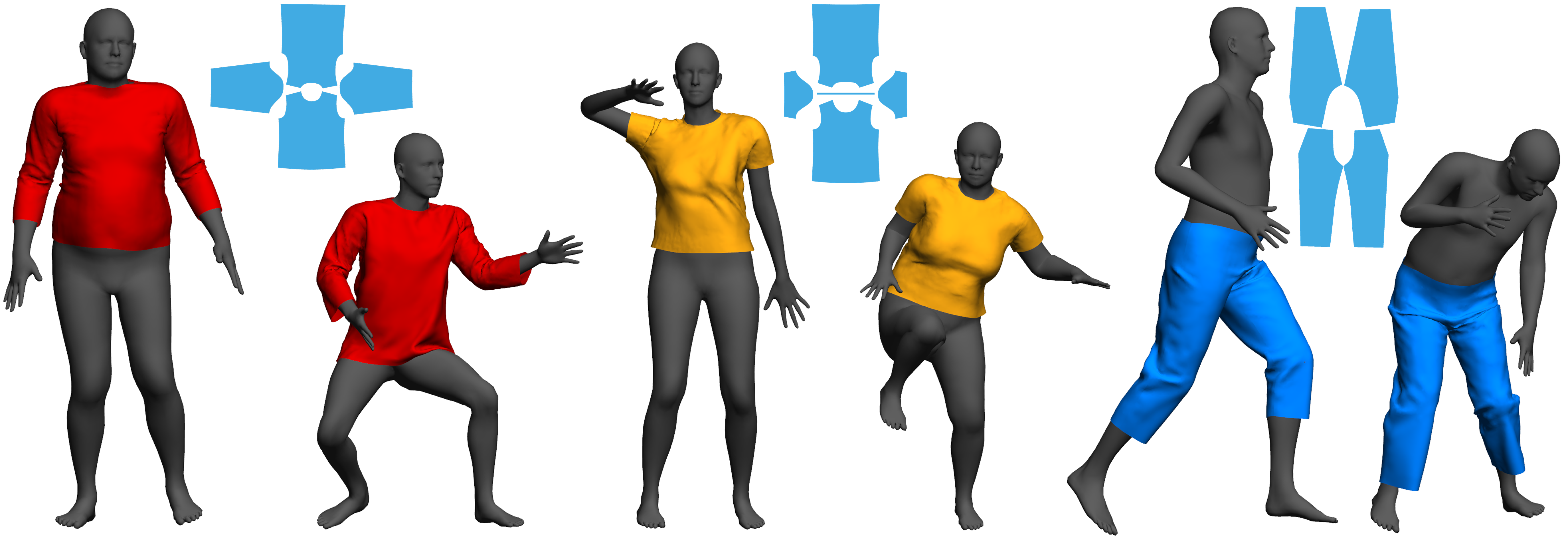}
  \caption{Synthesis results. Clothing patterns are shown in blue.}
\label{fig:moreresult}
\end{figure*}

 We use SMPL \cite{loper2015smpl} to generate template body meshes under different shapes and apply dual quaternion blending \cite{kavan2008geometric} to animate the body mesh. The skinning weights are calculated using Pinnochio \cite{baran2007automatic} and the motion sequences are from the CMU motion capture library \cite{cmumotiondata}.

 At run-time, given a new input pose, we employ Eq. \ref{eq:blending} to obtain a coarse result and then add the decaying effect and resolve interpenetrations to get the final result. Given a new input shape, we re-compute the new clothing instance $f(\beta, \theta_s)$ and update the binding information for each clothing shape model (See Sec. \ref{subsec:runtimeshythesis}), which takes 0.022s each time. In our implementation, all clothing shape models can be handled in parallel, and it takes 0.7s to handle 150 data points for the men's long-sleeved shirt.

Theoretically, our algorithm can add the shape dimension to any clothing pose model. To demonstrate this, we run our method with both SOR and a sequence of simulated clothing instances.
First, we use SOR as our clothing pose model. We turn off the penetration handling process in SOR and leave it for our method to address. We also discard the translation item (rightmost item in Eq. 1 of \cite{xu2014sensitivity}) in SOR since we find it has little impact on the final result, especially when using our sampling method. These measures contribute to the high speed of our method. We use our sampling method to generate both the SOR database and our database. As shown in Figure~\ref{fig:moreresult} and Figure~\ref{fig:teaser}, our method can predict clothing instances with fine wrinkles. Please refer to Table \ref{tbl:statistics} for detailed statistics.

\begin{figure*}[t]
\setlength{\abovecaptionskip}{0.0cm}
\centering
  \includegraphics[width=0.98\textwidth]{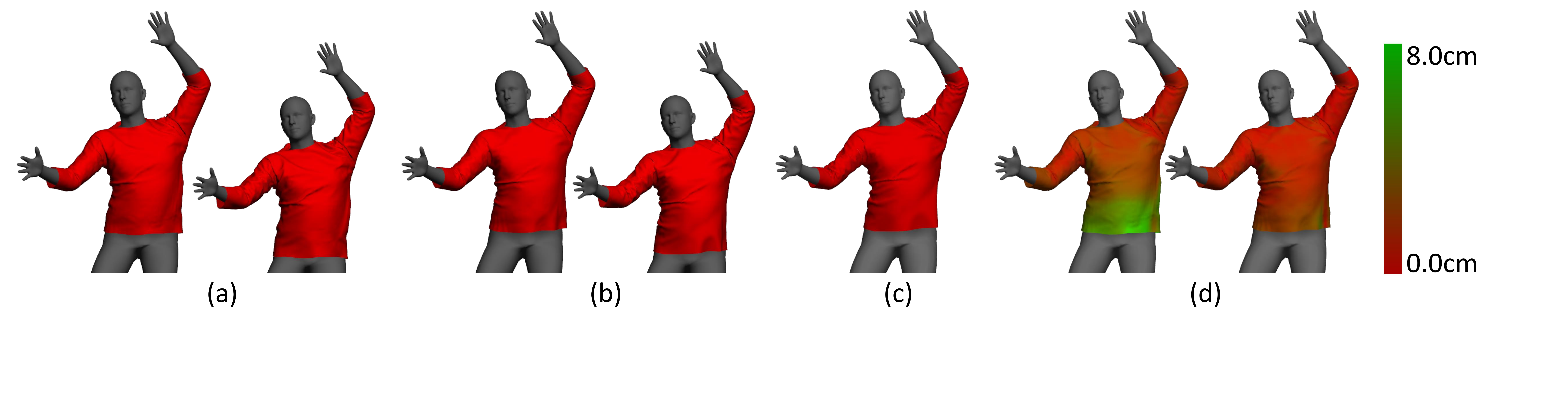}
  \caption{Comparison between the synthesis results achieved using SOR as the clothing pose model and those achieved using simulation as the clothing pose model. (a) Synthesis result (left) using SOR (right); (b) synthesis result (left) using simulation (right); (c) simulation result; (d) Euclidean vertex distance between the synthesis results and the simulation clothing. The errors are marked in green. We can see that the synthesis result achieved using simulation (right) has fewer errors than the result achieved using \purple{SOR (left)}.}
\label{fig:animationcomparison}
\end{figure*}

Second, we use simulated clothing instances as our clothing pose model. We simulate the garment mesh under a randomly chosen motion sequence while keeping the same body shape parameters. Then we use the resulting clothing instances as the clothing pose model in Eq. \ref{eq:blending}. Figure \ref{fig:animationcomparison}(b) shows that our method can recover realistic fold patterns.
We can see that the quality of our result relies on the external clothing pose model, i.e. if the clothing pose model is more accurate, our result is closer to the ground truth (Figure \ref{fig:animationcomparison}(c)). This can also be seen in Figure~\ref{fig:convergence}. We will elaborate on this in the following section.


\subsection{Convergence}
\begin{figure}[t]
\setlength{\abovecaptionskip}{0.0cm}
\centering
  \includegraphics[width=0.48\textwidth]{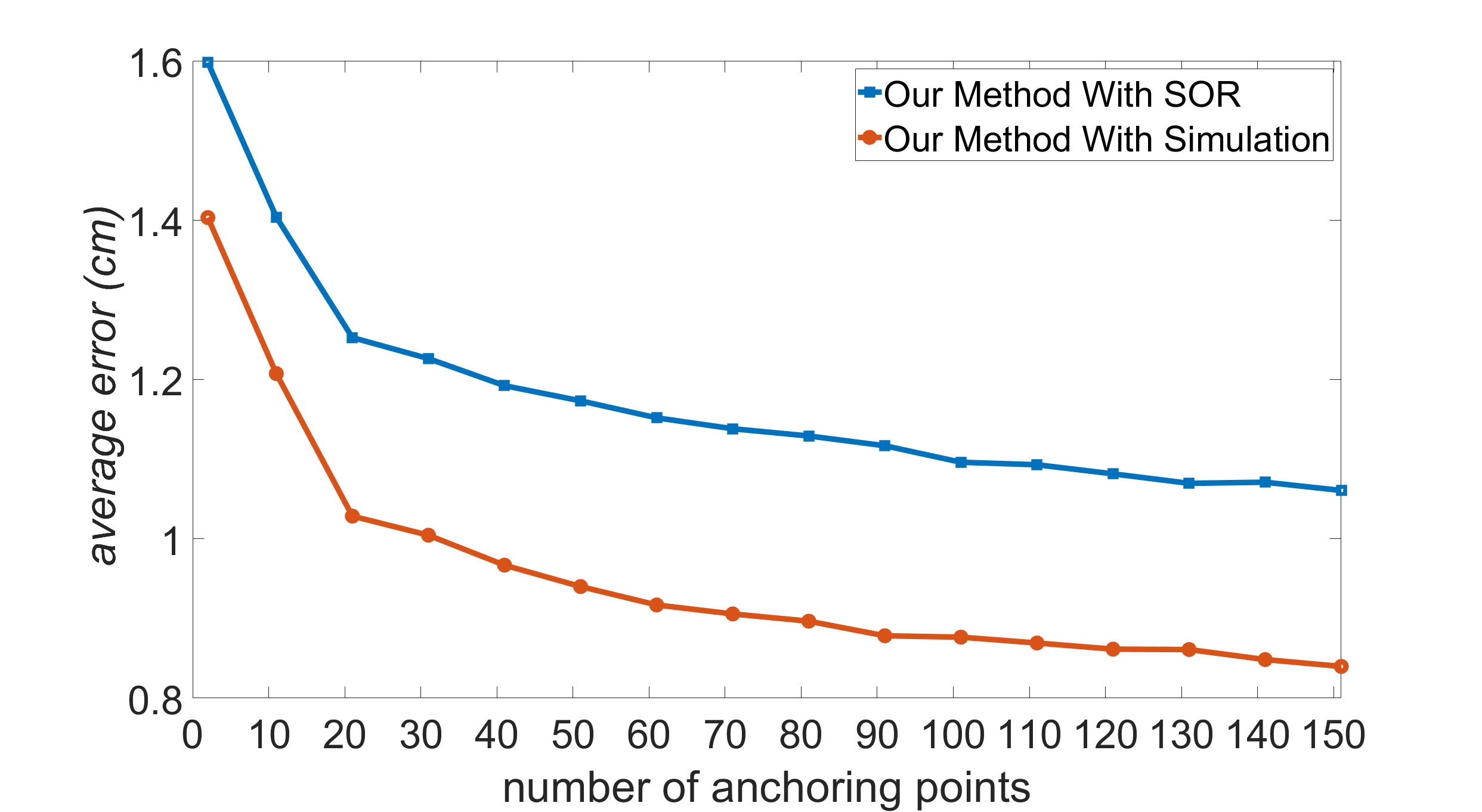}
  \caption{Convergence of synthesis error while using an increasing number of data points.}
\label{fig:convergence}
\end{figure}

We use the average Euclidean vertex distance between the synthesized clothing and the simulation clothing as the error to measure the physical accuracy of our result. The error is calculated over 16$\times$400 randomly chosen shape and pose pairs (16 shapes and 400 poses for each shape). As shown in Figure~\ref{fig:convergence}, the error of the \textit{results with SOR} drops to 1.06 cm at 150 data points, which we believe is acceptable in most scenarios. Given the number of data points, the error of the \textit{results with simulation} is much less than that of \textit{results with SOR}, which demonstrates that our method can be improved if a more accurate clothing pose model is employed.

\subsection{VR Scenarios}
\begin{figure}[t]
\setlength{\abovecaptionskip}{0.0cm}
\centering
  \includegraphics[width=0.475\textwidth]{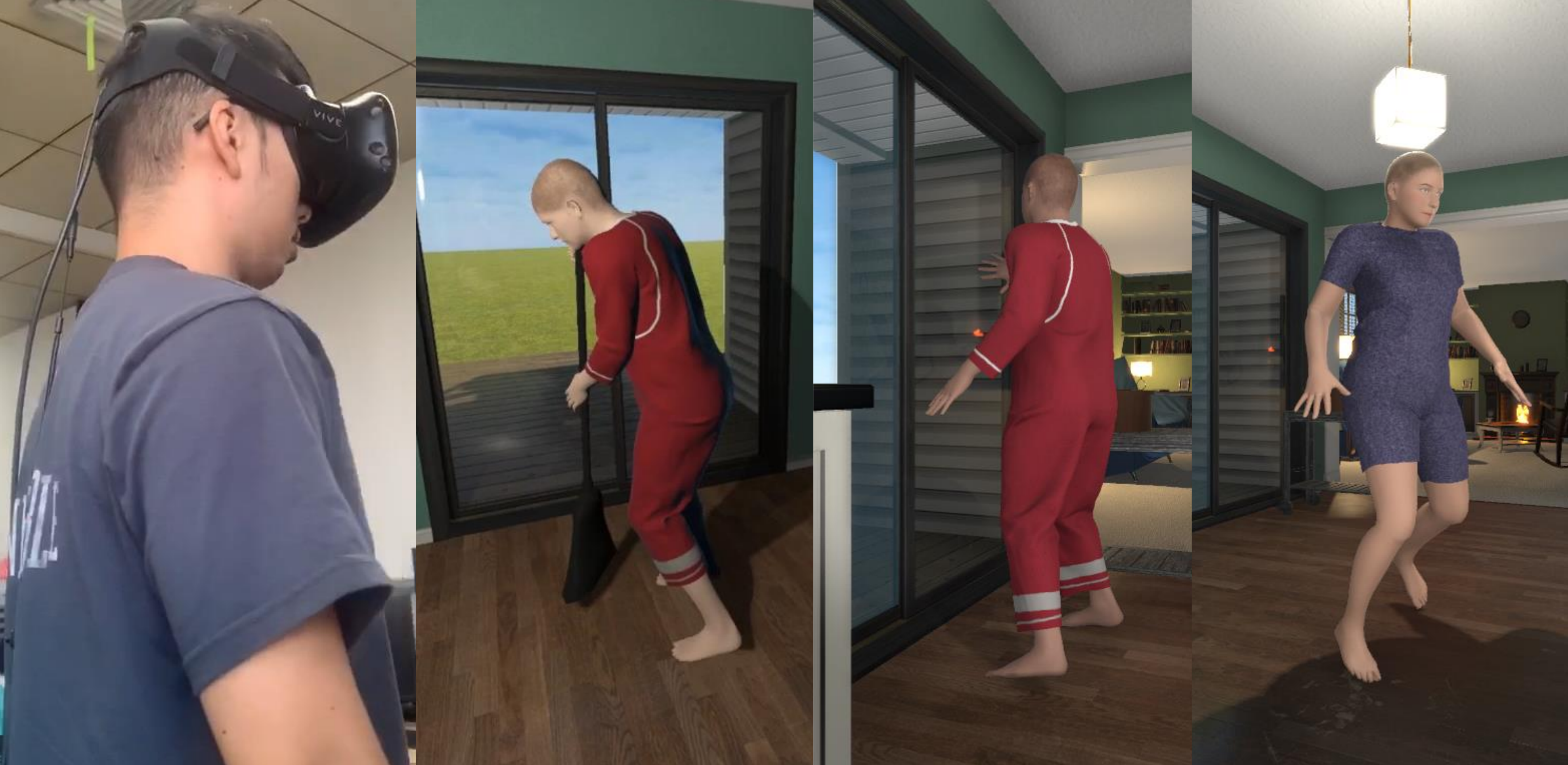}
  \caption{The avatar in a VR scenario. We provide the user with an immersive VR experience from a first-person perspective with HTC Vive. \purple{The avatar looks around and observes an agent performing some tasks.}}
\label{fig:vrapplications}
\end{figure}
Our method can be applied to VR scenarios. As shown in Figure \ref{fig:vrapplications},  we provide the user with an immersive VR experience from a first-person perspective with an HTC Vive (left most). The clothing deformations for the agents are generated by our method.

\subsection{User Evaluation}
\label{subsecuserstudy}

\begin{figure*}[t]
\setlength{\abovecaptionskip}{0.0cm}
\centering
  \includegraphics[width=1.0\textwidth]{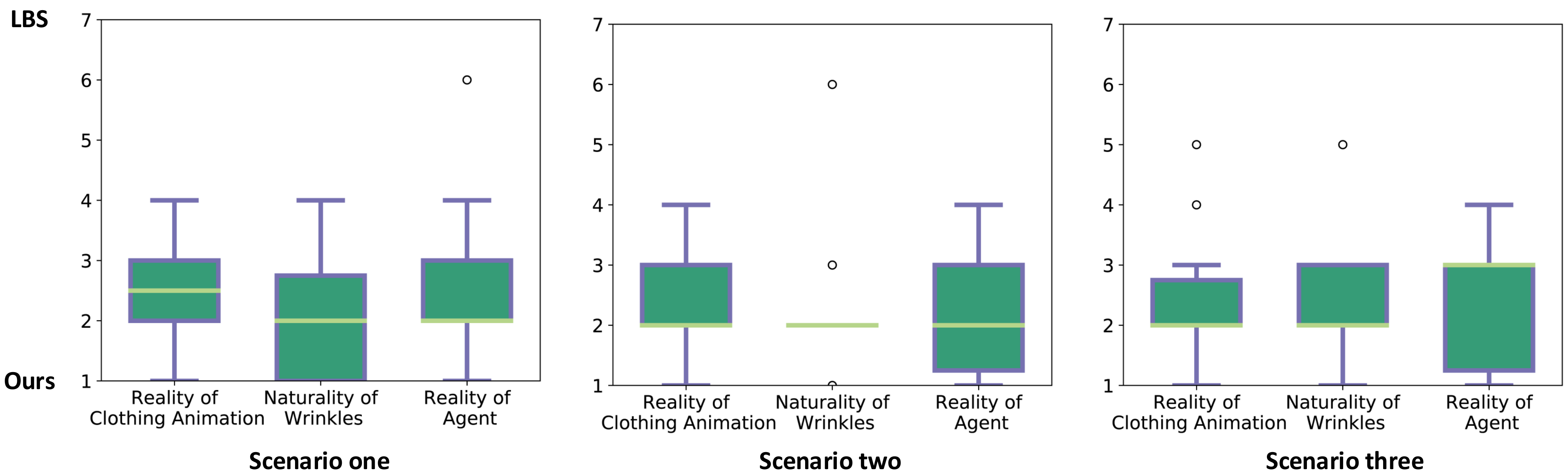}
  \caption{Participant preferences in user evaluation. The questions were ``Which clothing animation looks more realistic?'', ``Which wrinkles look more natural?'', and ``Which agent looks more like a real person?''. Scores are normalized such that 1 indicates a strong preference for our method and 7 indicates a strong preference for LBS method. \purple{For each scenario, participants rated our method preferably in terms of generating more realistic or plausible clothing animation ($2.44 \pm 1.042$, $2.28 \pm 0.895$, $2.22 \pm 1.060$), generating more natural-looking wrinkles ($2.17 \pm 1.150$, $2.06 \pm 1.110$, $2.50 \pm 1.043$), and enhancing the presence of the agent ($2.56 \pm 1.338$, $2.33 \pm 1.138$, $2.50 \pm 1.150$).}}
\label{fig:userstudystatistics}
\end{figure*}

\begin{table*}[htbp!]
\begin{center}
\resizebox{1.0\textwidth}{!}
{
\begin{tabular}{|l|c|c|c|c|c|c|c|c|c|c|c|c|c|c|}
\hline
Question (Which...) & 1 & 2  & 3  & 4  & 5  & 6  & 7 & mean & SD\\
\hline
clothing animation looks more realistic? & 4 / 4 / 4 & 5 / 6 / 9 & 6 / 7 / 3  & 3 / 1 / 1 & 0 / 0 / 1 & 0 / 0 / 0 & 0 / 0 / 0& 2.44 / 2.28 / 2.22 & $\pm 1.042$ / $\pm 0.895$ / $\pm 1.060$\\
\hline
wrinkles look more natural? & 6 / 4 / 1 & 7 / 12 / 11 & 1 / 1 / 4 & 4 / 0 / 0 & 0 / 0 / 2 & 0 / 1 /0 & 0 / 0 / 0 & 2.17 / 2.06 / 2.50 & $\pm 1.150$ / $\pm 1.110$ / $\pm 1.043$\\
\hline
agent looks more like a real person? & 4 / 5 / 5 & 6 / 6 / 3 & 4 / 3 / 6 & 3 / 4 / 4 & 0 / 0 / 0 & 1 / 0 / 0 & 0 / 0 / 0 & 2.56 / 2.33 / 2.50 & $\pm 1.338$ / $\pm 1.138$ / $\pm 1.150$\\
\hline
\end{tabular}
}
\end{center}
\caption{The response frequency of subjects in the study described in Sec. \ref{subsecuserstudy}. \purple{Scores are normalized such that $1$ indicates a strong preference for our method and $7$ indicates a strong preference for the LBS method. For each scenario (separated by ``/''), participants prefer our method over the prior method on several dimensions.}}
\label{tbl:userstudydata}
\end{table*}

We conducted a user study to demonstrate the perceptual benefit of our method compared to the prior technique in generating clothing deformations for the agent in immersive settings. 

\textbf{Experiment Goals \& Expectations:} We hypothesize that the clothing deformations generated by our method will exhibit more detailed wrinkles and more dynamics compared to prior methods and that participants will strongly prefer our results to those of prior methods.

\textbf{Experimental Design:} The study was conducted based on a within-subjects, paired-comparison design. Participants explored two simulations with a fixed exposure time wearing a HTC Vive headset. The clothing deformations in the two simulations were generated using our method and the LBS method. \purple{The order of scenes was counterbalanced, as well as the order of the methods.} After these two simulations, participants answered a set of questions, \purple{and our questionnaire design is inspired by prior methods~\cite{garau2005responses,slater2006analysis,narang2016pedvr}.}

\textbf{Comparison Methods:} Previous methods generally adopt skinning methods to deform the clothing mesh of an agent in virtual environments. Therefore, we evaluated our method against the Linear Blend Skinning (LBS) method. 

\textbf{Environments:} 
\purple{We use three scenarios for the user study. The first scenario was comprised of a man sweeping in the living room, while wearing a long-sleeved shirt and pants. The second scenario consisted of a man that has a different shape and wiping the windows, while wearing the same clothes as the first scenario. The last scenario corresponded to a woman wandering in a living room, while wearing a t-shirt and shorts. } Participants can walk around and observe the agent doing some tasks. Please refer to Figure \ref{fig:vrapplications} and the supplemental video for more details.

\textbf{Metrics:} Participants were asked to indicate their preference for a method using a 7-point Likert scale, with 1 indicating a strong preference for the method presented first, 7 indicating a strong preference for the method presented second, and 4 indicating no preference. \purple{In terms of reporting the results, we normalized the participant responses so that 1 indicates a strong preference for our method.}

\textbf{Results:} Our study was taken by 18 participants, 9 male, with a mean age of $24.44\pm 1.95$ years. The participant responses clearly demonstrate the benefits of our algorithm. For each question, we performed a one-sample $t$-test comparing the mean of the question with a hypothetical mean of 4 (no preference or no impact). The question ``Which clothing animation looks more realistic?'' was shown to be significant for each scenario: $t$(17) = -6.336, $p < 0.001$; $t$(17) = -8.166, $p < 0.001$; $t$(17) = -7.114, $p < 0.001$. The question ``Which wrinkles look more natural?'' was also significant for each scenario: $t$(17) = -6.761, $p < 0.001$; $t$(17) = -7.432, $p < 0.001$; $t$(17) = -6.101, $p < 0.001$, as was ``Which agent looks more like a real person?'': $t$(17) = -4.579, $p < 0.001$; $t$(17) = -6.216, $p < 0.001$; $t$(17) = -5.532, $p < 0.001$;. Figure
\ref{fig:userstudystatistics} and Table \ref{tbl:userstudydata} provide further details on participant responses.

\section{Discussion}\label{sec:discussion}
The rationale of our sampling method is that Taylor expansion locally approximates surrounding values of an anchoring point while cluster centers are the best set of points to approximate all the points, i.e. they are the best set of anchoring points. Compared to the MCMC sampling process used in SOR \cite{xu2014sensitivity}, our sampling process is able to sample more typical poses in real life, while their sampling step finds a pose that maximizes the error of the synthesized clothing, which might produce weird poses. Experimental results show that SOR results can be significantly improved when using our sampling method. Additionally, thanks to the mutual independence of our data points, our database can be constructed in parallel, while the MCMC process in SOR must run serially since their sampling process depends on the previously constructed data. Please refer to Sec. \ref{sec:experiment} for details.

{\textbf{Limitations.}}
Our method assumes that the clothing coordinate differences introduced by body shape variation are linear with body pose, i.e. the clothing coordinate differences caused by body shape variation under the new pose can be predicted from the anchoring pose using LBS. This hinders the application of our approach to garments that are loose on the avatar.

Furthermore, the final results of our method are highly dependent on the clothing pose model used. If the clothing pose model is not accurate, then our result is also not accurate. In addition, the efficiency of the clothing pose model creates a bottle-neck in our method. To overcome this difficulty, we plan to develop our own clothing pose model in the future.

The wrinkles and folds generated in the animation suffer from some sudden changes when body poses change too quickly. The reason for this is that our scheme is trained on clothing instances of static equilibrium, and the estimations from different poses are inconsistent. Although we have employed a decaying factor to smooth the animation, it cannot solve this problem  completely.

Finally, our method cannot guarantee that all the penetrations are resolved, especially when the clothing is too tight on the body. In this case, the LBS result for each anchoring pose may result in deep penetrations, and the blending result will make penetrations even worse, which is beyond the ability of our penetration resolving method to address. When this is the case, we believe that the clothing is not suitable for the body, suggesting that the clothing is a bad fit. In addition, when the clothing has too many folds, self-penetration might occur in our synthesis results.

{\textbf{Future works.}}
We want to extend our method to other control parameters like clothing material and clothing patterns. For clothing material, for example, we first need to devise a clothing material model. Given a set of clothing material parameters, this model can predict the corresponding clothing deformation. Then, we sample some valuable poses as the Taylor expansion anchoring points. Finally, we blend local approximation results at run-time.

 We currently use the average sensitivity of 17 training shapes to calculate the distances of anchoring points. However, as the body shape changes, the sensitivity changes accordingly. In the future, we plan to \purple{train} a sensitivity model, i.e. a model that can predict the corresponding sensitivity of the clothing given a new set of body shape parameters. We believe this will help us to find better anchoring points at run-time.


\section{Conclusion}\label{sec:conclusion}
In this paper, we have presented a clothing synthesis scheme that uses Taylor expansion to combine two independent components: the pose-dependent deformation and the shape-dependent deformation. As a result, our method can add the shape dimension to various clothing pose models. Our method does not need to modify or retrain the clothing pose model. The core innovation here is that we regard clothing deformation as a function of body shape parameters and body pose parameters and use Taylor expansion to locally approximate clothing deformations around anchoring points. Due to the high computational cost of higher-order derivatives, we use linear Taylor expansion in both offline and online processes. Our clothing shape model can efficiently predict realistic clothing deformations under various body shapes and has the potential for use as a stand-alone application.

Using only a CPU, our method can generate realistic clothing deformations under various body poses and shapes in real time without resizing the cloth model. Furthermore, we believe that our method can be extended to add other parameters such as clothing material and clothing pattern to clothing pose models. The performance can improve based on data-driven parameter estimation methods~\cite{wolinski2014parameter}.

%
%
%
%


%% file: spa_tvcg.bbl
\begin{thebibliography}{10}
\providecommand{\url}[1]{#1}
\csname url@samestyle\endcsname
\providecommand{\newblock}{\relax}
\providecommand{\bibinfo}[2]{#2}
\providecommand{\BIBentrySTDinterwordspacing}{\spaceskip=0pt\relax}
\providecommand{\BIBentryALTinterwordstretchfactor}{4}
\providecommand{\BIBentryALTinterwordspacing}{\spaceskip=\fontdimen2\font plus
\BIBentryALTinterwordstretchfactor\fontdimen3\font minus
  \fontdimen4\font\relax}
\providecommand{\BIBforeignlanguage}[2]{{%
\expandafter\ifx\csname l@#1\endcsname\relax
\typeout{** WARNING: IEEEtran.bst: No hyphenation pattern has been}%
\typeout{** loaded for the language `#1'. Using the pattern for}%
\typeout{** the default language instead.}%
\else
\language=\csname l@#1\endcsname
\fi
#2}}
\providecommand{\BIBdecl}{\relax}
\BIBdecl

\bibitem{bailenson2005independent}
J.~N. Bailenson, K.~Swinth, C.~Hoyt, S.~Persky, A.~Dimov, and J.~Blascovich,
  ``The independent and interactive effects of embodied-agent appearance and
  behavior on self-report, cognitive, and behavioral markers of copresence in
  immersive virtual environments,'' \emph{Presence: Teleoperators \& Virtual
  Environments}, vol.~14, no.~4, pp. 379--393, 2005.

\bibitem{baraff1998large}
D.~Baraff and A.~Witkin, ``Large steps in cloth simulation,'' in
  \emph{Proceedings of the 25th annual conference on Computer graphics and
  interactive techniques}.\hskip 1em plus 0.5em minus 0.4em\relax ACM, 1998,
  pp. 43--54.

\bibitem{narain2012adaptive}
R.~Narain, A.~Samii, and J.~F. O'Brien, ``Adaptive anisotropic remeshing for
  cloth simulation,'' \emph{ACM Trans. Graph.}, vol.~31, no.~6, p. 152, 2012.

\bibitem{cirio2014yarn}
G.~Cirio, J.~Lopez-Moreno, D.~Miraut, and M.~A. Otaduy, ``Yarn-level simulation
  of woven cloth,'' \emph{ACM Trans. Graph.}, vol.~33, no.~6, p. 207, 2014.

\bibitem{kim2013near}
D.~Kim, W.~Koh, R.~Narain, K.~Fatahalian, A.~Treuille, and J.~F. O'Brien,
  ``Near-exhaustive precomputation of secondary cloth effects,'' \emph{ACM
  Trans. Graph.}, vol.~32, no.~4, p.~87, 2013.

\bibitem{xu2014sensitivity}
W.~Xu, N.~Umentani, Q.~Chao, J.~Mao, X.~Jin, and X.~Tong,
  ``Sensitivity-optimized rigging for example-based real-time clothing
  synthesis,'' \emph{ACM Trans. Graph.}, vol.~33, no.~4, p. 107, 2014.

\bibitem{guan2012drape}
P.~Guan, L.~Reiss, D.~A. Hirshberg, A.~Weiss, and M.~J. Black, ``Drape:
  Dressing any person.'' \emph{ACM Trans. Graph.}, vol.~31, no.~4, pp. 35--1,
  2012.

\bibitem{santesteban2019learning}
I.~Santesteban, M.~A. Otaduy, and D.~Casas, ``Learning-based animation of
  clothing for virtual try-on,'' \emph{arXiv preprint arXiv:1903.07190}, 2019.

\bibitem{hauth2003analysis}
M.~Hauth, O.~Etzmu{\ss}, and W.~Stra{\ss}er, ``Analysis of numerical methods
  for the simulation of deformable models,'' \emph{The Visual Computer},
  vol.~19, no. 7-8, pp. 581--600, 2003.

\bibitem{volino2005implicit}
P.~Volino and N.~Magnenat-Thalmann, ``Implicit midpoint integration and
  adaptive damping for efficient cloth simulation,'' \emph{Computer Animation
  and Virtual Worlds}, vol.~16, no. 3-4, pp. 163--175, 2005.

\bibitem{fierz2011element}
B.~Fierz, J.~Spillmann, and M.~Harders, ``Element-wise mixed implicit-explicit
  integration for stable dynamic simulation of deformable objects,'' in
  \emph{Proceedings of the 2011 ACM SIGGRAPH/Eurographics Symposium on Computer
  Animation}.\hskip 1em plus 0.5em minus 0.4em\relax ACM, 2011, pp. 257--266.

\bibitem{provot1995deformation}
X.~Provot \emph{et~al.}, ``Deformation constraints in a mass-spring model to
  describe rigid cloth behaviour,'' in \emph{Graphics interface}.\hskip 1em
  plus 0.5em minus 0.4em\relax Canadian Information Processing Society, 1995,
  pp. 147--147.

\bibitem{goldenthal2007efficient}
R.~Goldenthal, D.~Harmon, R.~Fattal, M.~Bercovier, and E.~Grinspun, ``Efficient
  simulation of inextensible cloth,'' \emph{ACM Trans. Graph.}, vol.~26, no.~3,
  p.~49, 2007.

\bibitem{thomaszewski2009continuum}
B.~Thomaszewski, S.~Pabst, and W.~Strasser, ``Continuum-based strain
  limiting,'' in \emph{Computer Graphics Forum}, vol.~28, no.~2.\hskip 1em plus
  0.5em minus 0.4em\relax Wiley Online Library, 2009, pp. 569--576.

\bibitem{wang2010multi}
H.~Wang, J.~O'Brien, and R.~Ramamoorthi, ``Multi-resolution isotropic strain
  limiting,'' in \emph{ACM Trans. Graph.}, vol.~29, no.~6.\hskip 1em plus 0.5em
  minus 0.4em\relax ACM, 2010, p. 156.

\bibitem{ma2016anisotropic}
G.~Ma, J.~Ye, J.~Li, and X.~Zhang, ``Anisotropic strain limiting for
  quadrilateral and triangular cloth meshes,'' in \emph{Computer Graphics
  Forum}, vol.~35, no.~1.\hskip 1em plus 0.5em minus 0.4em\relax Wiley Online
  Library, 2016, pp. 89--99.

\bibitem{grinspun2003discrete}
E.~Grinspun, A.~N. Hirani, M.~Desbrun, and P.~Schr{\"o}der, ``Discrete
  shells,'' in \emph{Proceedings of the 2003 ACM SIGGRAPH/Eurographics
  symposium on Computer animation}.\hskip 1em plus 0.5em minus 0.4em\relax
  Eurographics Association, 2003, pp. 62--67.

\bibitem{english2008animating}
E.~English and R.~Bridson, ``Animating developable surfaces using nonconforming
  elements,'' in \emph{ACM Trans. Graph.}, vol.~27, no.~3.\hskip 1em plus 0.5em
  minus 0.4em\relax ACM, 2008, p.~66.

\bibitem{choi2005stable}
K.-J. Choi and H.-S. Ko, ``Stable but responsive cloth,'' in \emph{ACM SIGGRAPH
  2005 Courses}.\hskip 1em plus 0.5em minus 0.4em\relax ACM, 2005, p.~1.

\bibitem{volino2009simple}
P.~Volino, N.~Magnenat-Thalmann, and F.~Faure, ``A simple approach to nonlinear
  tensile stiffness for accurate cloth simulation,'' \emph{ACM Trans. Graph.},
  vol.~28, no.~4, pp. Article--No, 2009.

\bibitem{bridson2002robust}
R.~Bridson, R.~Fedkiw, and J.~Anderson, ``Robust treatment of collisions,
  contact and friction for cloth animation,'' in \emph{ACM Trans. Graph.},
  vol.~21, no.~3.\hskip 1em plus 0.5em minus 0.4em\relax ACM, 2002, pp.
  594--603.

\bibitem{harmon2008robust}
D.~Harmon, E.~Vouga, R.~Tamstorf, and E.~Grinspun, ``Robust treatment of
  simultaneous collisions,'' \emph{ACM Trans. Graph.}, vol.~27, no.~3, p.~23,
  2008.

\bibitem{zheng2012energy}
C.~Zheng and D.~L. James, ``Energy-based self-collision culling for arbitrary
  mesh deformations,'' \emph{ACM Trans. Graph.}, vol.~31, no.~4, p.~98, 2012.

\bibitem{guo2018material}
Q.~Guo, X.~Han, C.~Fu, T.~Gast, R.~Tamstorf, and J.~Teran, ``A material point
  method for thin shells with frictional contact,'' \emph{ACM Trans. Graph.},
  vol.~37, no.~4, p. 147, 2018.

\bibitem{liu2013fast}
T.~Liu, A.~W. Bargteil, J.~F. O'Brien, and L.~Kavan, ``Fast simulation of
  mass-spring systems,'' \emph{ACM Trans. Graph.}, vol.~32, no.~6, p. 214,
  2013.

\bibitem{bouaziz2014projective}
S.~Bouaziz, S.~Martin, T.~Liu, L.~Kavan, and M.~Pauly, ``Projective dynamics:
  fusing constraint projections for fast simulation,'' \emph{ACM Trans.
  Graph.}, vol.~33, no.~4, p. 154, 2014.

\bibitem{wang2015chebyshev}
H.~Wang, ``A chebyshev semi-iterative approach for accelerating projective and
  position-based dynamics,'' \emph{ACM Trans. Graph.}, vol.~34, no.~6, p. 246,
  2015.

\bibitem{fratarcangeli2018parallel}
M.~Fratarcangeli, H.~Wang, and Y.~Yang, ``Parallel iterative solvers for
  real-time elastic deformations,'' in \emph{SIGGRAPH Asia 2018 Courses}, 2018,
  pp. 1--45.

\bibitem{muller2007position}
M.~M{\"u}ller, B.~Heidelberger, M.~Hennix, and J.~Ratcliff, ``Position based
  dynamics,'' \emph{Journal of Visual Communication and Image Representation},
  vol.~18, no.~2, pp. 109--118, 2007.

\bibitem{tang2016cama}
M.~Tang, H.~Wang, L.~Tang, R.~Tong, and D.~Manocha, ``Cama: Contact-aware
  matrix assembly with unified collision handling for gpu-based cloth
  simulation,'' in \emph{Computer Graphics Forum}, vol.~35, no.~2.\hskip 1em
  plus 0.5em minus 0.4em\relax Wiley Online Library, 2016, pp. 511--521.

\bibitem{tang2018cloth}
M.~Tang, T.~Wang, Z.~Liu, R.~Tong, and D.~Manocha, ``I-cloth: incremental
  collision handling for gpu-based interactive cloth simulation,'' \emph{ACM
  Transactions on Graphics (TOG)}, vol.~37, no.~6, pp. 1--10, 2018.

\bibitem{tang2013gpu}
M.~Tang, R.~Tong, R.~Narain, C.~Meng, and D.~Manocha, ``A gpu-based streaming
  algorithm for high-resolution cloth simulation,'' in \emph{Computer Graphics
  Forum}, vol.~32, no.~7.\hskip 1em plus 0.5em minus 0.4em\relax Wiley Online
  Library, 2013, pp. 21--30.

\bibitem{govindaraju2005quick}
N.~K. Govindaraju, M.~C. Lin, and D.~Manocha, ``Quick-cullide: Fast inter-and
  intra-object collision culling using graphics hardware,'' in \emph{IEEE
  Proceedings. VR 2005. Virtual Reality, 2005.}\hskip 1em plus 0.5em minus
  0.4em\relax IEEE, 2005, pp. 59--66.

\bibitem{pcloth}
C.~Li, M.~Tang, R.~Tong, M.~Cai, J.~Zhao, and D.~Manocha, ``P-cloth:
  interactive complex cloth simulation on multi-gpu systems using dynamic
  matrix assembly and pipelined implicit integrators,'' \emph{ACM Transactions
  on Graphics (TOG)}, vol.~39, no.~6, pp. 1--15, 2020.

\bibitem{manocha1994algorithms}
D.~Manocha and J.~Demmel, ``Algorithms for intersecting parametric and
  algebraic curves i: simple intersections,'' \emph{ACM Transactions on
  Graphics (TOG)}, vol.~13, no.~1, pp. 73--100, 1994.

\bibitem{tang2014fast}
M.~Tang, R.~Tong, Z.~Wang, and D.~Manocha, ``Fast and exact continuous
  collision detection with bernstein sign classification,'' \emph{ACM
  Transactions on Graphics (TOG)}, vol.~33, no.~6, pp. 1--8, 2014.

\bibitem{manocha1998solving}
D.~Manocha, ``Solving polynomial equations,'' \emph{Applications of
  Computational Algebraic Geometry: American Mathematical Society Short Course,
  January 6-7, 1997, San Diego, California}, vol.~53, p.~41, 1998.

\bibitem{manocha1992algebraic}
------, \emph{Algebraic and numeric techniques in modeling and robotics}.\hskip
  1em plus 0.5em minus 0.4em\relax University of California at Berkeley, 1992.

\bibitem{klosowski1998efficient}
J.~T. Klosowski, M.~Held, J.~S.~B. Mitchell, H.~Sowizral, and K.~Zikan,
  ``Efficient collision detection using bounding volume hierarchies of
  k-dops,'' \emph{IEEE Transactions on Visualization and Computer Graphics},
  vol.~4, no.~1, pp. 21--36, Jan. 1998.

\bibitem{fuhrmann2003distance}
A.~Fuhrmann, G.~Sobotka, and C.~Gro{\ss}, ``Distance fields for rapid collision
  detection in physically based modeling,'' in \emph{Proceedings of GraphiCon
  2003}.\hskip 1em plus 0.5em minus 0.4em\relax Citeseer, 2003, pp. 58--65.

\bibitem{thiery2013sphere}
J.-M. Thiery, {\'E}.~Guy, and T.~Boubekeur, ``Sphere-meshes: Shape
  approximation using spherical quadric error metrics,'' \emph{ACM Trans.
  Graph.}, vol.~32, no.~6, p. 178, 2013.

\bibitem{wu2018variational}
N.~Wu, D.~Zhang, Z.~Deng, and X.~Jin, ``Variational mannequin approximation
  using spheres and capsules,'' \emph{IEEE Access}, vol.~6, pp.
  25\,921--25\,929, 2018.

\bibitem{teschner2005collision}
M.~Teschner, S.~Kimmerle, B.~Heidelberger, G.~Zachmann, L.~Raghupathi,
  A.~Fuhrmann, M.-P. Cani, F.~Faure, N.~Magnenat-Thalmann, W.~Strasser, and
  P.~Volino, ``{Collision Detection for Deformable Objects},'' in
  \emph{Eurographics 2004 - STARs}.\hskip 1em plus 0.5em minus 0.4em\relax
  Eurographics Association, 2004.

\bibitem{de2010stable}
E.~De~Aguiar, L.~Sigal, A.~Treuille, and J.~K. Hodgins, ``Stable spaces for
  real-time clothing,'' in \emph{ACM Trans. Graph.}, vol.~29, no.~4.\hskip 1em
  plus 0.5em minus 0.4em\relax ACM, 2010, p. 106.

\bibitem{wang2010example}
H.~Wang, F.~Hecht, R.~Ramamoorthi, and J.~F. O'Brien, ``Example-based wrinkle
  synthesis for clothing animation,'' in \emph{ACM Trans. Graph.}, vol.~29,
  no.~4.\hskip 1em plus 0.5em minus 0.4em\relax ACM, 2010, p. 107.

\bibitem{hahn2014subspace}
F.~Hahn, B.~Thomaszewski, S.~Coros, R.~W. Sumner, F.~Cole, M.~Meyer, T.~DeRose,
  and M.~Gross, ``Subspace clothing simulation using adaptive bases,''
  \emph{ACM Trans. Graph.}, vol.~33, no.~4, p. 105, 2014.

\bibitem{jin2018pixel}
N.~Jin, Y.~Zhu, Z.~Geng, and R.~Fedkiw, ``A pixel-based framework for
  data-driven clothing,'' \emph{arXiv preprint arXiv:1812.01677}, 2018.

\bibitem{chentanez2020cloth}
N.~Chentanez, M.~Macklin, M.~M{\"u}ller, S.~Jeschke, and T.-Y. Kim, ``Cloth and
  skin deformation with a triangle mesh based convolutional neural network,''
  in \emph{Computer Graphics Forum}, vol.~39, no.~8.\hskip 1em plus 0.5em minus
  0.4em\relax Wiley Online Library, 2020, pp. 123--134.

\bibitem{anguelov2005scape}
D.~Anguelov, P.~Srinivasan, D.~Koller, S.~Thrun, J.~Rodgers, and J.~Davis,
  ``Scape: shape completion and animation of people,'' in \emph{ACM Trans.
  Graph.}, vol.~24, no.~3.\hskip 1em plus 0.5em minus 0.4em\relax ACM, 2005,
  pp. 408--416.

\bibitem{wang2018learning}
T.~Y. Wang, D.~Ceylan, J.~Popovic, and N.~J. Mitra, ``Learning a shared shape
  space for multimodal garment design,'' \emph{arXiv preprint
  arXiv:1806.11335}, 2018.

\bibitem{loper2015smpl}
M.~Loper, N.~Mahmood, J.~Romero, G.~Pons-Moll, and M.~J. Black, ``Smpl: A
  skinned multi-person linear model,'' \emph{ACM Trans. Graph.)}, vol.~34,
  no.~6, p. 248, 2015.

\bibitem{lahner2018deepwrinkles}
Z.~Lahner, D.~Cremers, and T.~Tung, ``Deepwrinkles: Accurate and realistic
  clothing modeling,'' in \emph{Proceedings of the European Conference on
  Computer Vision (ECCV)}, 2018, pp. 667--684.

\bibitem{patel2020tailornet}
C.~Patel, Z.~Liao, and G.~Pons-Moll, ``Tailornet: Predicting clothing in 3d as
  a function of human pose, shape and garment style,'' in \emph{Proceedings of
  the IEEE/CVF Conference on Computer Vision and Pattern Recognition}, 2020,
  pp. 7365--7375.

\bibitem{ma2020learning}
Q.~Ma, J.~Yang, A.~Ranjan, S.~Pujades, G.~Pons-Moll, S.~Tang, and M.~J. Black,
  ``Learning to dress 3d people in generative clothing,'' in \emph{Proceedings
  of the IEEE/CVF Conference on Computer Vision and Pattern Recognition}, 2020,
  pp. 6469--6478.

\bibitem{yang2018analyzing}
J.~Yang, J.-S. Franco, F.~H{\'e}troy-Wheeler, and S.~Wuhrer, ``Analyzing
  clothing layer deformation statistics of 3d human motions,'' in
  \emph{Proceedings of the European Conference on Computer Vision (ECCV)},
  2018, pp. 237--253.

\bibitem{saito2019pifu}
S.~Saito, Z.~Huang, R.~Natsume, S.~Morishima, A.~Kanazawa, and H.~Li, ``Pifu:
  Pixel-aligned implicit function for high-resolution clothed human
  digitization,'' in \emph{Proceedings of the IEEE International Conference on
  Computer Vision}, 2019, pp. 2304--2314.

\bibitem{natsume2019siclope}
R.~Natsume, S.~Saito, Z.~Huang, W.~Chen, C.~Ma, H.~Li, and S.~Morishima,
  ``Siclope: Silhouette-based clothed people,'' in \emph{Proceedings of the
  IEEE Conference on Computer Vision and Pattern Recognition}, 2019, pp.
  4480--4490.

\bibitem{pons2017clothcap}
G.~Pons-Moll, S.~Pujades, S.~Hu, and M.~J. Black, ``Clothcap: Seamless 4d
  clothing capture and retargeting,'' \emph{ACM Transactions on Graphics
  (TOG)}, vol.~36, no.~4, pp. 1--15, 2017.

\bibitem{yang2017learning}
S.~Yang, J.~Liang, and M.~C. Lin, ``Learning-based cloth material recovery from
  video,'' in \emph{Proceedings of the IEEE International Conference on
  Computer Vision}, 2017, pp. 4383--4393.

\bibitem{saltelli2004sensitivity}
A.~Saltelli, S.~Tarantola, F.~Campolongo, and M.~Ratto, ``Sensitivity analysis
  in practice: a guide to assessing scientific models,'' \emph{Chichester,
  England}, 2004.

\bibitem{kiendl2014isogeometric}
J.~Kiendl, R.~Schmidt, R.~W{\"u}chner, and K.-U. Bletzinger, ``Isogeometric
  shape optimization of shells using semi-analytical sensitivity analysis and
  sensitivity weighting,'' \emph{Computer Methods in Applied Mechanics and
  Engineering}, vol. 274, pp. 148--167, 2014.

\bibitem{zehnder2017metasilicone}
J.~Zehnder, E.~Knoop, M.~B{\"a}cher, and B.~Thomaszewski, ``Metasilicone:
  design and fabrication of composite silicone with desired mechanical
  properties,'' \emph{ACM Trans. Graph.}, vol.~36, no.~6, p. 240, 2017.

\bibitem{geilinger2018skaterbots}
M.~Geilinger, R.~Poranne, R.~Desai, B.~Thomaszewski, and S.~Coros,
  ``Skaterbots: Optimization-based design and motion synthesis for robotic
  creatures with legs and wheels,'' \emph{ACM Trans. Graph.}, vol.~37, no.~4,
  p. 160, 2018.

\bibitem{zimmermann2019puppetmaster}
S.~Zimmermann, R.~Poranne, J.~M. Bern, and S.~Coros, ``Puppetmaster: robotic
  animation of marionettes,'' \emph{ACM Trans. Graph.}, vol.~38, no.~4, p. 103,
  2019.

\bibitem{umetani2011sensitive}
N.~Umetani, D.~M. Kaufman, T.~Igarashi, and E.~Grinspun, ``Sensitive couture
  for interactive garment modeling and editing.'' \emph{ACM Trans. Graph.},
  vol.~30, no.~4, pp. 90--1, 2011.

\bibitem{barrielle2019realtime}
V.~Barrielle and N.~Stoiber, ``Realtime performance-driven physical simulation
  for facial animation,'' in \emph{Computer Graphics Forum}, vol.~38,
  no.~1.\hskip 1em plus 0.5em minus 0.4em\relax Wiley Online Library, 2019, pp.
  151--166.

\bibitem{shen2015geometrically}
Z.~Shen, J.~Huang, W.~Chen, and H.~Bao, ``Geometrically exact simulation of
  inextensible ribbon,'' in \emph{Computer Graphics Forum}, vol.~34,
  no.~7.\hskip 1em plus 0.5em minus 0.4em\relax Wiley Online Library, 2015, pp.
  145--154.

\bibitem{chen2010fully}
M.~Chen and K.~Tang, ``A fully geometric approach for developable cloth
  deformation simulation,'' \emph{The visual computer}, vol.~26, no. 6-8, pp.
  853--863, 2010.

\bibitem{bergou2007tracks}
M.~Bergou, S.~Mathur, M.~Wardetzky, and E.~Grinspun, ``Tracks: toward
  directable thin shells,'' \emph{ACM Trans. Graph.}, vol.~26, no.~3, pp.
  50--es, 2007.

\bibitem{cmumotiondata}
CMU, ``{CMU} graphics lab motion capture database.''
  \emph{http://mocap.cs.cmu.edu.}, 2003.

\bibitem{kavan2008geometric}
L.~Kavan, S.~Collins, J.~{\v{Z}}{\'a}ra, and C.~O'Sullivan, ``Geometric
  skinning with approximate dual quaternion blending,'' \emph{ACM Trans.
  Graph.}, vol.~27, no.~4, p. 105, 2008.

\bibitem{baran2007automatic}
I.~Baran and J.~Popovi{\'c}, ``Automatic rigging and animation of 3d
  characters,'' in \emph{ACM Trans. Graph.}, vol.~26, no.~3.\hskip 1em plus
  0.5em minus 0.4em\relax ACM, 2007, p.~72.

\bibitem{garau2005responses}
M.~Garau, M.~Slater, D.-P. Pertaub, and S.~Razzaque, ``The responses of people
  to virtual humans in an immersive virtual environment,'' \emph{Presence:
  Teleoperators \& Virtual Environments}, vol.~14, no.~1, pp. 104--116, 2005.

\bibitem{slater2006analysis}
M.~Slater, C.~Guger, G.~Edlinger, R.~Leeb, G.~Pfurtscheller, A.~Antley,
  M.~Garau, A.~Brogni, and D.~Friedman, ``Analysis of physiological responses
  to a social situation in an immersive virtual environment,'' \emph{Presence:
  Teleoperators and Virtual Environments}, vol.~15, no.~5, pp. 553--569, 2006.

\bibitem{narang2016pedvr}
S.~Narang, A.~Best, T.~Randhavane, A.~Shapiro, and D.~Manocha, ``Pedvr:
  Simulating gaze-based interactions between a real user and virtual crowds,''
  in \emph{Proceedings of the 22nd ACM conference on virtual reality software
  and technology}, 2016, pp. 91--100.

\bibitem{wolinski2014parameter}
D.~Wolinski, S.~J.~Guy, A.-H. Olivier, M.~Lin, D.~Manocha, and J.~Pettr{\'e},
  ``Parameter estimation and comparative evaluation of crowd simulations,'' in
  \emph{Computer Graphics Forum}, vol.~33, no.~2.\hskip 1em plus 0.5em minus
  0.4em\relax Wiley Online Library, 2014, pp. 303--312.

\end{thebibliography}
